\documentclass{article}
\usepackage{ijcai22}

\usepackage{times}
\usepackage{soul}
\usepackage{url}
\usepackage[hidelinks]{hyperref}
\usepackage[utf8]{inputenc}
\usepackage[small]{caption}
\usepackage{subcaption}
\usepackage{graphicx}
\usepackage{amsmath}
\usepackage{amssymb}
\usepackage{amsthm}
\usepackage{multirow,makecell}
\usepackage{booktabs}
\usepackage{pgf, tikz}
\usepackage{bm}
\usepackage{dsfont}
\usepackage{mathtools}
\usepackage{float}
\usepackage{filecontents}
\usepackage{standalone}
\usepackage{algorithm}
\usepackage{algorithmic}
\newcommand{\X}{\mathcal{X}}

\newcommand{\sipomdplitenettitle}{\textsf{sIPOMDPLite-Net}}
\newcommand{\sipomdplitenettext}{\textsf{sIPOMDPLite-net}}
\def\FreeGold{\makecell[l]{$s_i\in$FREESPACE\\ $s_j=$GOLD}}
\def\GoldGold{\makecell[l]{$s_i=$GOLD\\ $s_j=$GOLD}}
\def\FreeMinusGoldGold{\makecell[l]{$s_i\in$FREESPACE$\setminus$GOLD\\ $s_j=$GOLD}}
\def\primeObstFree{\makecell[l]{$s_i^{\prime}\in$OBSTACLES\\ $s_i^{\prime}\in$\underline{FREESPACE}}}
\def\BmNonprimePrime{\makecell[l]{$\bm{s_i^{\prime}}\in s_i$\\ $\bm{s_i^{\prime}}\in s_i^{\prime}$}} 
\def\BmPrimeFree{\makecell[l]{$\bm{s_i^{\prime}}\in$\underline{FREESPACE}\\ $\bm{s_i^{\prime}}\in$\underline{FREESPACE}}}

\graphicspath{{figs/}}
\tikzset{every picture/.style={line width=0.75pt}} 

\title{\sipomdplitenettitle{}: Lightweight, Self-Interested Learning and Planning in POSGs with Sparse Interactions}

\author{
Gengyu Zhang
\and
Prashant Doshi
\affiliations
University of Georgia, Athens, USA\\
\{gengyu.zhang, pdoshi\}@uga.com,
}

\begin{document}

\maketitle

\begin{abstract}
  This work introduces \sipomdplitenettext{}, a deep neural network (DNN) architecture for decentralized, self-interested agent control in partially observable stochastic games (POSGs) with sparse interactions between agents. The network learns to plan in contexts modeled by the interactive partially observable Markov decision process (I-POMDP) Lite framework and uses hierarchical value iteration networks to simulate the solution of nested MDPs, which I-POMDP Lite attributes to the other agent to model its behavior and predict its intention. We train \sipomdplitenettext{} with expert demonstrations on small two-agent Tiger-grid tasks, for which it accurately learns the underlying I-POMDP Lite model and near-optimal policy, and the policy continues to perform well on larger grids and real-world maps. As such, \sipomdplitenettext{} shows good transfer capabilities and offers a lighter learning and planning approach for individual, self-interested agents in multiagent settings.
\end{abstract}

\section{Introduction}
\label{sec:intro}

Recent deep reinforcement learning (DRL) methods~\cite{game-theoretic-DMARL-meanfield,game-theoretic-DMARL-grand-SC2} that aim to solve POSGs are mostly model-free, which derive agents' optimal policies by computing the Nash equilibria~\cite{game-theoretic-framework:DP-POSGs,game-theoretic-framework:MARL}. However, such methods are inexplicable and typically data-inefficient.

In contrast, a model-based vein of investigations introduces NN analogs of well-founded decision-theoretic planning frameworks and makes them learn underlying models such as the Markovian transition and observation function from data. This vein has succeeded in the field of single-agent learning and planning~\cite{neural-learning-planning:VIN,neural-learning-planning:QMDP-net}. Nevertheless, extending the approach to multiagent settings is challenging. For instance, \cite{neural-learning-planning:IPOMDP-net} introduced an NN analog of the self-interested decision-theoretic framework, I-POMDP~\cite{decision-theoretic-framework-solver:apprx-I-PF,decision-theoretic-framework-solver:apprx-I-PBVI,critique:I-POMDP}. To alleviate the exacerbated curses of dimensionality and history~\cite{decision-theoretic-framework:I-POMDP-Lite} in solving the I-POMDP, they introduced the I-PF and QMDP algorithm into their NN architecture. However, it is not clear how they made the sampling-based operations differentiable for the end-to-end training\footnote{The source code of IPOMDP-net is unavailable, so we do not know its implementational details of the network architecture and experiments.}.

To tackle the issues from the ground up, we continue with this vein and propose an NN architecture based on the more pragmatic I-POMDP Lite~\cite{decision-theoretic-framework:I-POMDP-Lite} framework. First, the framework eliminates the curse of dimensionality by assuming other agents have perfect observability and thus embedding a nested MDP to model their behaviors. Then, we employ the multiagent QMDP~\cite{qmdp-1,qmdp-2} (MAQMDP) to solve the I-POMDP Lite, which mitigates the curse of history. Furthermore, we introduce the sparse interaction~\cite{Dec-SIMDP,heuristic-Dec-SIMDP} to offset the enormous joint state-action space, where the models reduce to single-agent ones in non-interactive situations.

The \sipomdplitenettext{} represents a policy for a class of multiagent decision-making tasks. It encodes the I-POMDP Lite models into the NN architecture and trains it supervised by the expert demonstration. Our experiments train the network on trajectory data for a set of relatively small grids and show that the learned NN model continues to plan well when the agents are situated in larger grids and a realistically complex environment. The empirical results indicate that our network successfully learns the underlying logic of the problem, which helps to predict the other agent's behavior and plan optimal actions accordingly for the subjective agent.


\section{I-POMDP Lite Framework}
\label{sec:I-POMDP-Lite}
In this section, we introduce the I-POMDP Lite~\cite{decision-theoretic-framework:I-POMDP-Lite}, a lightweight self-interested multiagent planning framework that lays the theoretical foundation for \sipomdplitenettext{}.

Considering two self-interested agents, $i$ and $j$, in a multiagent system, we define an I-POMDP Lite model from $i$'s perspective as $\langle b_0, S, A, \Omega_i, T_i, O_i, R_i, \hat{\pi}_j, \gamma\rangle$. $b_0$ is agent $i$'s initial belief over common physical states, $S$, of the two agents; $\Omega_i$ is the set of local observations of agent $i$; $A=A_i\times A_j$ is the set of joint actions. The definition of the transition function, $T_i$, reward function, $R_i$, and observation function, $O_i$, is similar to that of a POMDP, except that joint actions in the multiagent context now determine them. $\hat{\pi}_j$: $S\times A_j\to[0,1]$ is the predicted mixed strategy of agent $j$, indicating $i$'s belief about each $a_j\in A_j$ selected in each $s\in S$, i.e., $\Pr(a_j|s)$. $\gamma\in(0,1)$ is a discounted factor.

The I-POMDP Lite employs a nested MDP framework to solve for $\hat{\pi}_j$. Assuming that $j$ reasons at level $l^{\prime}$, the top-level of the nested structure is thus $l^{\prime}$, and $\hat{\pi}_j$ can be further denoted as $\hat{\pi}_j^{l^{\prime}}$. We formally define a nested MDP that models $j$ as $M_j^{l^{\prime}} = \langle S,A,T_j,R_j,\{\pi_i^l\}_{l=0}^{l^{\prime}-1},\gamma\rangle$. $\{\pi_i^l\}_{l=0}^{l^{\prime}-1}$: $S\times A_i\to[0,1]$ is a set of $i$'s policies at each nested reasoning level $l$, where $0\leq l<l^{\prime}$. The optimal $(k+1)$-step-to-go value function of $M_j^l$ for agent $j$ reasoning at level $l^{\prime}$ satisfies the following Bellman equation.
\begin{align}
  \label{eq:nested-mdp-vi}
  \begin{aligned}
    {Q_j^{l^{\prime},k+1}(s,a)} &= R_j(s,a) + \gamma\sum_{s^{\prime}}T_j(s,a,s^{\prime})\\
    &\max_{a_j^{\prime}}\sum_{a_i^{\prime}}\hat{\pi}_i^{l^{\prime}}(s^{\prime},a_i^{\prime})Q_j^{l^{\prime},k}(s^{\prime},a^{\prime})
  \end{aligned}
\end{align}

When $l^{\prime}=0$, agent $j$'s mixed strategy $\hat{\pi}_j^0$ is simply $|A_j|^{-1}$, which uniformly selects $a_j$ in whatever state; otherwise, there is $\hat{\pi}_j^{l^{\prime}}(s,a_j) = \sum_{l=0}^{l^{\prime}}\Pr(l)\pi_j^l(s,a_j)$, where $\Pr(l)$ is the probability of which agent $j$ reasons at level $l\in [0,l^{\prime}]$, and each $\pi_j^l$ is derived by solving the corresponding $M_j^l$.

Provided with the full observability of nested MDPs, agent $i$ need not model agent $j$'s belief as in I-POMDP. Instead, its belief dimensionality remains $S\times A_j$. We can further factorize it to separate out the belief over physical states only, where $b_i(s,a_j) = b_i(s)\hat{\pi}_j(s,a_j)$. Hence, we essentially transfer the model to a POMDP variant that maintains the other agent's policy prediction.  Eq.~\ref{eq:ipomdplite-bu} provides the I-POMDP Lite belief update. $\eta$ denotes the normalization operation.
\begin{align}
    \label{eq:ipomdplite-bu}
    \begin{aligned}
      b_i^{\prime}(s^{\prime}) = \eta O_i(s^{\prime}, a_i, o_i^{\prime})\sum_{s,a_j}T_i(s, a, s^{\prime})\Pr(a_j|s)b_i(s)
    \end{aligned}
\end{align}

\begin{figure*}[tp]
  \centering
  \includegraphics[width=0.8\textwidth]{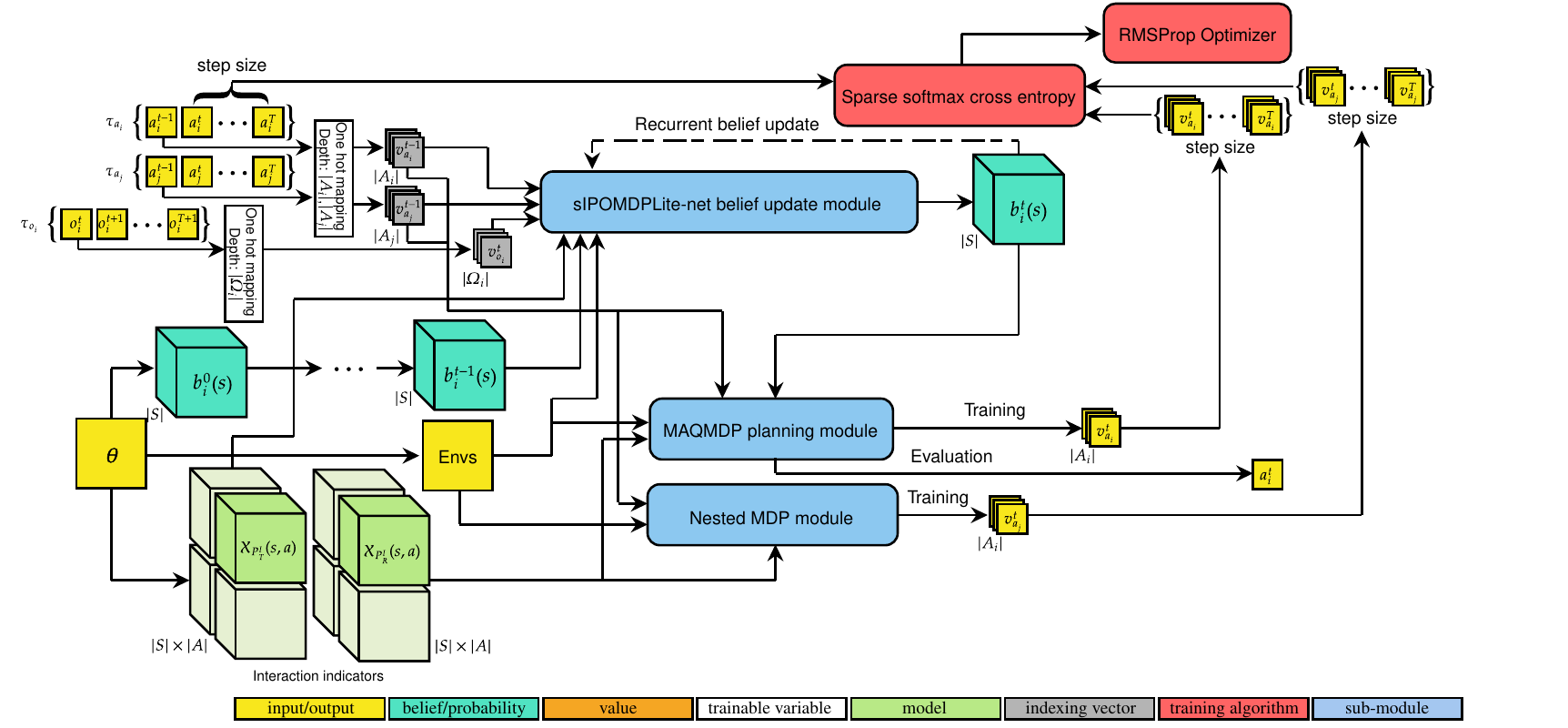}
  \caption{The general \sipomdplitenettext{} architecture, consisting of the top-level I-POMDP Lite reasoning and planning module and the nested MDP module, maps the current joint action and observation sampled from an expert's demonstration and the task parameter to the subjective agent's next action. The network is trained by minimizing the cross-entropy between the output and the expert's action trajectories.}
  \label{fig:network_arch_toplevel}
\end{figure*}

\section{sIPOMDPLite-Net}
\label{sec:network}

In this section, we elucidate the \sipomdplitenettext{} architecture based on the I-POMDP Lite framework and sparse interactions. We first introduce the macroscopic architecture, followed by the factorization of the original model under sparse interactions in mathematics and the demonstration of each major component's architecture. For ease of discussion, in the rest of the section, we assume only two agents -- the subjective agent $i$ and the objective agent $j$. Besides, the nested MDP only has one reasoning level.

\subsection{Overview}
\label{subsec:network:overview}

The \sipomdplitenettext{} represents a policy of a class of tasks modeled by the I-POMDP Lite. For any agent $i$ and agent $j$ in such a task, the network maps the provided task features, agents' current joint action, $a=\langle a_i,a_j\rangle$, and agent $i$'s local observation, $o_i$, to its next action to take, $a_i^{\prime}$, which we denote by $\pi_i(a_i,a_j,o_i,\bm{\theta}) = a_i^{\prime}$.

We introduce the task parameter $\bm{\theta}$ to integrate and parameterize the task features that determine the agents' transitions, rewards, and observations in a task. The set of all such parameters, denoted by $\bm{\Theta}$, covers all combinations of possible values of a wide variety of task features that vary within the class of tasks, and each $\bm{\theta}$ unambiguously identifies a task. Each $\bm{\theta}$ in our work comprises the initial belief, environment-related data, and the interaction indicator function of both agents.

The other input is the expert's joint action and observation trajectories. We generate them by solving the I-POMDP Lite models for the set of training tasks, where the resultant policy is regarded as the expert policy. Then, we simulate the agents' behaviors based on the expert policy in the task environments, storing the joint action and observation for each time step to frame the expert trajectories.

As Fig.~\ref{fig:network_arch_toplevel} illustrates, the \sipomdplitenettext{} iteratively takes joint actions and observations from the trajectories. The nested MDP module and MAQMDP module both receive a joint action at a time. They share an embedded recurrent architecture accounting for the value iteration, which maps task parameters to agents' $Q$ values. Then, the nested MDP module maps the $Q$ value of the joint input action to the softmax policy. On the other hand, the MAQMDP includes the belief in its policy search, which first weights the agent $i$'s joint action value by the current belief and then maps it to the softmax policy. The belief update module is a Bayesian filter recursively propagating the belief, which takes a tuple of joint action and observation at a time. The updated belief for each time step is sent to the MAQMDP planning module for computing the policy.

If the agents in a non-cooperative multiagent system interact sparsely, they only need to consider each other's influence in a small group of state-action pairs. Otherwise, they reference the private model, which is single-agent.

We formally define a set of boolean functions to indicate the interactive points as in Eq.~\ref{eq:def-ma-indicators}. If we use $P$, where $P=S\times A$, to denote the set of all $\langle s,a\rangle$ pairs, then $P^I$ is a subset in which the pairs pinpoint all potential interactive cases.
\begin{align}
    \label{eq:def-ma-indicators}
    \begin{aligned}
      \X_{P^I}(s,a) = 
      \begin{cases}
          1 & \text{if}~(s,a)\in P^I\\
          0 & \text{otherwise}
      \end{cases}\\
    \end{aligned}
\end{align}

In this paper, the indicator functions include the transition interaction indicator, $\X_{P^I_T}$, and the reward interaction indicator, $\X_{P^I_R}$. Typically, the transition and reward interactions are identical, i.e., $P_T^I = P_R^I$, and thus, $\X_{P^I_T} = \X_{P^I_R}$.

\subsection{Nested MDP Module}
\label{subsubsec:network:nested_mdp}

We decompose agent $i$'s reward and transition function with the interaction indicators as:
\begin{align}
    \label{eq:R_factorize}
    \begin{aligned}
      R_i(s,a) &= \Big[1 - \X_{P^I_R}(s,a)\Big]\mathfrak{r}_i(s_i,a_i) + \X_{P^I_R}(s,a)R_i(s,a)\\
      R_j(s,a) &= \Big[1 - \X_{P^I_R}(s,a)\Big]\mathfrak{r}_j(s_j,a_j) + \X_{P^I_R}(s,a)R_j(s,a)
    \end{aligned}
\end{align}
\begin{align}
    \label{eq:T_factorize}
    \begin{aligned}
      T_i(s,a,s^{\prime}) &= \Big[1 - \X_{P^I_T}(s,a)\Big]\mathfrak{t}_i(s_i,a_i,s_i^{\prime})\mathfrak{t}_j(s_j,a_j,s_j^{\prime})\\
      &+ \X_{P^I_T}(s,a)T_i(s,a,s^{\prime})
    \end{aligned}
\end{align}
where $R_i$ and $T_i$ are the original multiagent models and $\mathfrak{r}_i$, $\mathfrak{r}_j$, $\mathfrak{t}_i$, and $\mathfrak{t}_j$ are single-agent models only accounting for non-interactive situations.

Therefore, we rewrite the nested MDP value function as Eq.~\ref{eq:deriv-factorize-bellman}. Note that the non-interactive update of the expected reward is a two-step operation, where we first update the state utility by $\mathfrak{t}_j$ and then by $\mathfrak{t}_i$.
\begin{align}
    \label{eq:deriv-factorize-bellman}
    \begin{aligned}
     Q_j^{k+1}(s,a) &= \Big[1 - \X_{P^I_R}(s,a)\Big]\mathfrak{r}_j(s_j,a_j) + \X_{P^I_R}(s,a)R_j(s,a)\\
     &+ \gamma\Bigg\{\Big[1 - \X_{P^I_T}(s,a)\Big]\sum_{s_i^{\prime}}\mathfrak{t}_i(s_i,a_i,s_i^{\prime})\\
     &\sum_{s_j^{\prime}}\mathfrak{t}_j(s_j,a_j,s_j^{\prime})\max_{a_j}\sum_{a_i^{\prime}}\Pr(a_i^{\prime}|s^{\prime})Q_j^k(s^{\prime},a^{\prime})\\
     &+ \X_{P^I_T}(s,a)\sum_{s^{\prime}}T_j(s,a,s^{\prime})\max_{a_j}\sum_{a_i^{\prime}}\Pr(a_i^{\prime}|s^{\prime})\\
     &Q_j^k(s^{\prime},a^{\prime})\Bigg\}
    \end{aligned}
\end{align}

\begin{figure*}[tp]
  \centering
  \includegraphics[width=0.8\textwidth]{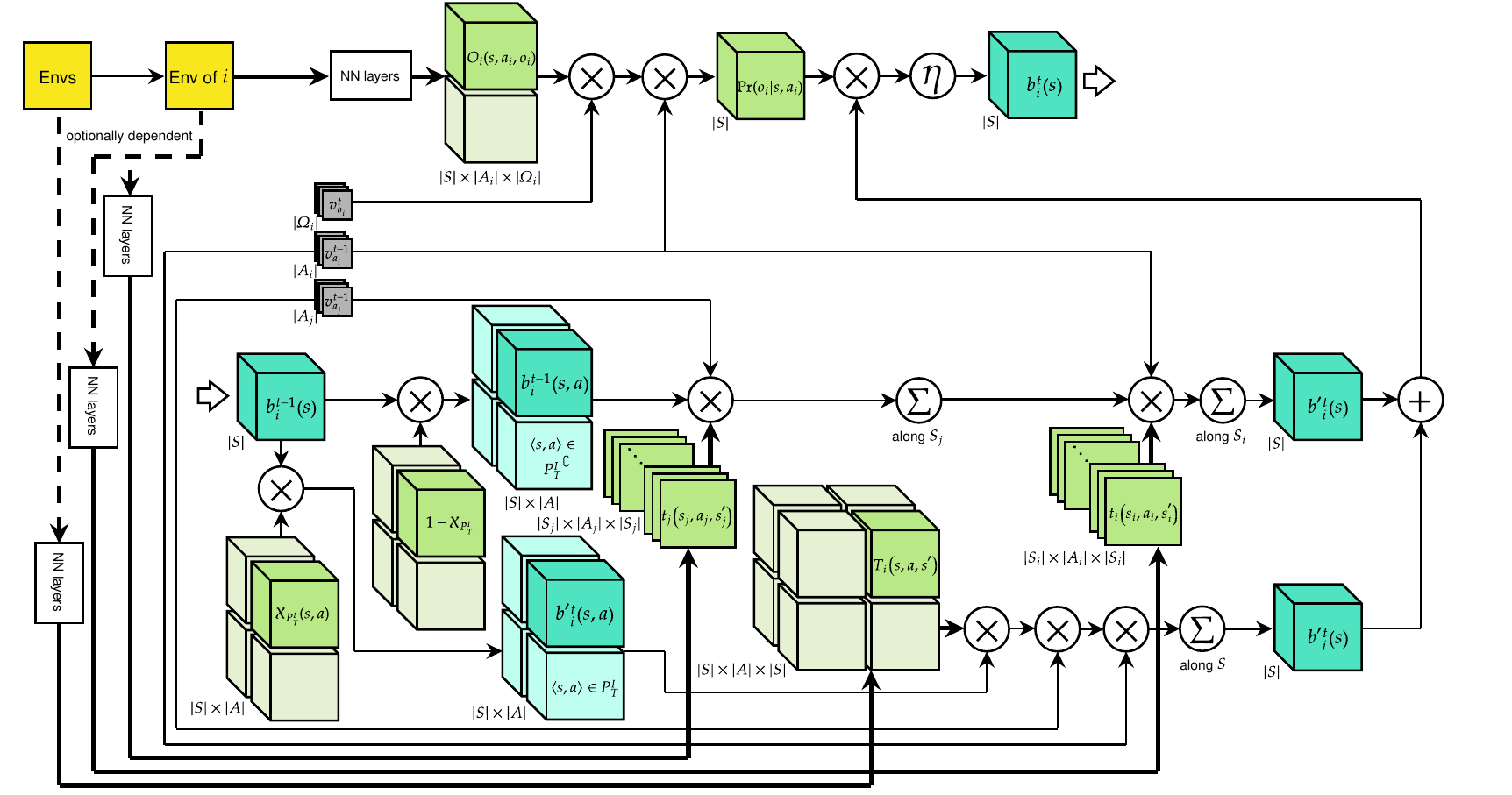}
  \caption{The belief update module of the \sipomdplitenettext{} is an RNN that recursively update the subjective agent's belief at each time step. The transition and observation model is learned through NN from the training task environments.}
  \label{fig:network_arch_beliefupdate}
\end{figure*}

\begin{figure*}[t]
  \centering
  \includegraphics[width=0.75\textwidth]{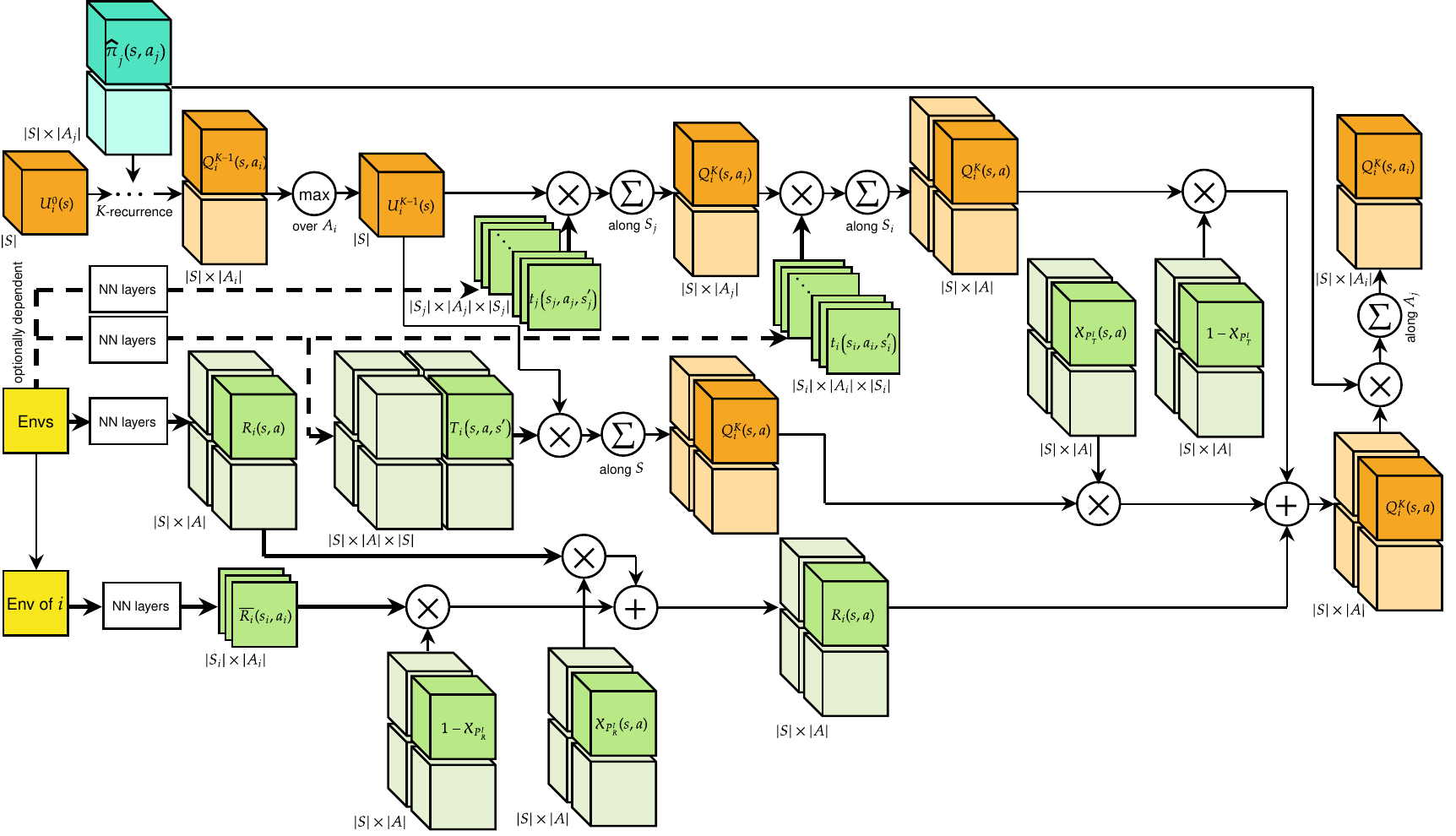}
  \caption{The value iteration architecture shared by the nested MDP and MAQMDP planning module is an RNN with $K$ recursions, where the hidden state is the state utility tensor. The transition and reward model is learned through NN from the training task environments.}
  \label{fig:network_arch_vi-solver}
\end{figure*}

The \sipomdplitenettext{} conditions the underlying Markovian model on the task parameter, $\bm{\theta}$, where a network architecture extracts related features from $\bm{\theta}$ and maps them to an approximate model. The specific class of networks to select depends on the task features. For instance, we demonstrate the use of convolutional NN (CNN) architecture to tackle tasks with the spatial locality in Appendix~\ref{subsec:appendix:implement}. Note that in this paper, we include the interaction indicator functions, $\X_{P^I_T}$ and $\X_{P^I_R}$, in task parameters as prior knowledge instead of learning through the network.

Agent $j$ reasons at the highest level of the nested MDP framework, so in the NN architecture, the model used to construct this level includes $T_j(\cdot|\bm{\theta})$, $R_j(\cdot|\bm{\theta})$, $\mathfrak{t}_i(\cdot|\bm{\theta})$, $\mathfrak{t}_j(\cdot|\bm{\theta})$, and $\mathfrak{r}_j(\cdot|\bm{\theta})$. The nested policy is given by the output of the lower-level architecture corresponding to agent $i$'s reasoning.

The nested MDP module comprises a value iteration solver and a softmax policy mapper for each level of the nested architecture. In the value iteration solver, we initialize the state utilities for the horizon length $0$, i.e., $U_j^0(s)$, as a zero tensor of the shape $|S|$. Then, based on Eq.~\ref{eq:deriv-factorize-bellman}, the tensor is updated by $T_j(\cdot|\bm{\theta})$ for $\langle s,a\rangle\in P^I$ while by $\mathfrak{t}_i(\cdot|\bm{\theta})$ and $\mathfrak{t}_j(\cdot|\bm{\theta})$ for $\langle s,a\rangle\in {P^I}^{\complement}$. Then, we filter the results, two $Q$ tensors, with $\bm{\X}_{P^I_T}$ and $1-\bm{\X}_{P^I_T}$, respectively, to ensure that both tensors have correct values. Next, we sum up the two $Q$ tensors and add the result to $R_j(\cdot|\bm{\theta})$, where we get $Q_j^1(s,a)$. The nested MDP module then weights $Q_j^1(s,a)$ by a pre-known policy, $\hat{\pi}_i$, to get $Q_j^1(s,a_j)$, the $Q$ values regarding only $a_j\in A_j$. Finally, we obtain $U_j^1(s)$ by picking the maximum over $Q_j^1(s,a_j)$, the input of the following recursion.

The recurrence continues for $K$ times, where $K$ is a general horizon selected for a set of tasks with the same scale, which is provided as a hyperparameter. Finally, we map the final $Q$ value to a softmax policy that indicates the best $a_i$ to select at the time step.

\subsection{Belief Update Module}
\label{subsec:network:beliefupdate}
Analogous to the nested MDP's value function, we factorize the original belief update as:
\begin{align}
    \label{eq:deriv-factorize-bu}
    \begin{aligned}
     b_i^{\prime}(s^{\prime}) &= \eta O_i(s^{\prime},a_i,o_i^{\prime})\Bigg\{\sum_{s_i}\mathfrak{t}_i(s_i,a_i,s_i^{\prime})\\
     &\sum_{s_j,a_j}\Big[1 - \X_{P^I_T}(s_j,a_j)\Big]\mathfrak{t}_j(s_j,a_j,s_j^{\prime})\Pr(a_j|s)b_i(s)\\
     &+ \sum_{s,a_j}\X_{P^I_T}(s,a)T_i(s,a,s^{\prime})\Pr(a_j|s)b_i(s)\Bigg\}
    \end{aligned}
\end{align}

We assume that either agent's observations are unaffected by the other's actions given the post-transition states. Hence, we can separate the observation function from the belief propagation enclosed in the curly brace.

The belief propagation works similarly to the expected value update in Eq.~\ref{eq:deriv-factorize-bellman}, except that in the belief propagation, we forwardly derive $b^{\prime}(s^{\prime})$ given $b(s)$, while in the value function, we backwardly deduce $U^{k+1}(s)$ from $U^k(s^{\prime})$. Thus, the indicator in Eq.~\ref{eq:deriv-factorize-bu} is placed inside the summation over the current state $s$ while that in Eq.~\ref{eq:deriv-factorize-bellman} is out of the summation over the next state $s^{\prime}$.

This time, the network conditions the transition and observation functions on $\bm{\theta}$. The transition functions represent the same underlying model as those in the nested MDP module, but we do not share their weights.

We depict the general architecture of the Bayesian filter in \ref{fig:network_arch_beliefupdate}. The input belief tensor is initially decomposed into the interactive and non-interactive part by the transition interaction indicator. Then, they are updated by the multiagent transitions, $T_i(\cdot|\bm{\theta})$, and the two-step single-agent transition, $\mathfrak{t}_j(\cdot|\bm{\theta})$ and $\mathfrak{t}_i(\cdot|\bm{\theta})$, respectively. Different from the framework, the belief update module of our network explicitly receives a joint action, including the other agent's action, instead of maintaining a predicted policy of the opponent inside the module. Thus, we eliminate the uncertainty from the nested MDP module, which ensures a good model learning of the top-level architecture. Next, the two separately updated beliefs are integrated and corrected by the observation, where the input $a_i$ and $o_i$ play a role as the index to match the accordant posterior probability distribution, $\Pr(o_i|s,a_i)$, from $O_i(\cdot|\bm{\theta})$. Finally, after an element-wise multiplication and normalization, we complete a step of the \sipomdplitenettext{} belief update.

\subsection{MAQMDP Planning Module}
\label{subsec:network:maqmdp}

The QMDP algorithm solves POMDPs approximately by replacing the original value function with the MDP value function, eliminating belief update in every step of the value iteration. The ultimately derived $Q$ values are weighted by the updated belief of each recurrence to represent the agent's updated preference for the actions. While sacrificing limited optimality, the algorithm provides a much faster and less expensive solution.

In this paper, we generalize the QMDP algorithm to multiagent settings. Coincidentally, solving I-POMDP Lite models can be regarded as solving POMDPs, except that it is aware of the other agent's intentions. Thus, the multiagent MAQMDP's value function is essentially equal to the nested MDP's as shown in Eq.~\ref{eq:nested-mdp-vi}. The MAQMDP inherits the QMDP's advantages over the original I-POMDP Lite solution, which makes our NN architecture lighter, avoiding the nested belief update module and thus benefiting the training.

Similar to the QMDP, we compute the value for each $a_i$ by weighting its corresponding $Q$ value with the updated belief for the next time step:
\begin{align}
    \label{eq:qmdp_policy}
    \begin{aligned}
     q_i(a_i) = \sum_{s,a_j}Q_i^K(s,a)\Pr(a_j|s)b_i(s)
    \end{aligned}
\end{align}
Finally, we can obtain the softmax policy based on $q_i(a_i)$.

Back to the \sipomdplitenettext{} architecture, the MAQMDP planning module shares most of its architecture with the nested MDP levels that solve agent $i$'s model. After the value iteration solver outputs the $Q$ function, we weight it by the belief tensor output by the belief update module for each recurrence and output the softmax policy to the trainer, where the cross-entropy between the output policy and the expert policy is used as the loss for training.

\section{Experiments}

We train and evaluate the \sipomdplitenettext{} on manually crafted Tiger-grids and their real-world extension. In this section, we first describe the setup of our experimental scenarios. Then, we briefly demonstrate the general training procedures of the network. Finally, we exhibit and interpret the results. In addition, we conduct a set of ablation studies to analyze the role that certain architectures play and elucidate the results in Appendix~\ref{subsec:appendix:ablation}.

\begin{figure*}[t]
  \centering
  \begin{subfigure}[t]{0.25\textwidth}
    \centering
    \includestandalone{tiger-grid-init}
    \caption{Initial state.}
    \label{fig:exp-tiger-grid-init}
  \end{subfigure}
  \hspace{0.0125\textwidth}
  \begin{subfigure}[t]{0.25\textwidth}
    \centering
    \includestandalone{tiger-grid-reset}
    \caption{Stochastic state reset.}
    \label{fig:exp-tiger-grid-reset}
  \end{subfigure}
  \hspace{0.0125\textwidth}
  \begin{subfigure}[t]{0.25\textwidth}
    \centering
    \includegraphics{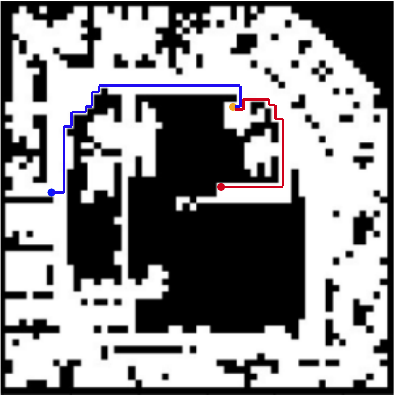}
    \caption{LIDAR maps.}
    \label{fig:exp-laser_maps}
  \end{subfigure}
  \caption{Examples of two experimental domains. (\subref{fig:exp-tiger-grid-init}) The initial state of the task, where the subjective agent does not know exactly the locations of others and itself but maintains a belief. (\subref{fig:exp-tiger-grid-reset}) A reset happening after an agent successfully seizes gold, where the system shifts each agent to a neighboring cell in either cardinal or intercardinal directions. (\subref{fig:exp-laser_maps}) A down-sampled real-world LIDAR map can be regarded as an extension of the Tiger-grids, in which we mainly evaluate the network's ability of generalization.}
  \label{fig:exp-tiger-grid}
\end{figure*}

\subsection{Experimental Setup}
\label{subsec:expr:setup}

Our experiments consist of two types of scenarios -- small synthetic two-agent Tiger grids and huge grids transferred from 2D LIDAR maps of realistic buildings. Agents in both types of tasks share the same actions, local transition properties, and rewards. This subsection focuses on the setup of agents' behaviors. Details of the underlying model including transition probabilities and exact rewards can be found in Appendix~\ref{subsec:appendix:setup}.

\subsubsection{Two-Agent Tiger-Grid Problem}
We introduce the Tiger-grid problem, a generalization of the classic Tiger problem to 2D grid world. Two robots act in a grid with randomly distributed obstacles. A door exists in each free cell, but only one door has a pile of gold behind it. We present such an environment in Fig.~\ref{fig:exp-tiger-grid-init}. The state is the raveled index of the cells where the robots occupy and the actions available for them are listening, moving, and opening the door in any free cell. While listening, a robot stays still and receives an observation truly reflecting the underlying state. Otherwise, it receives random observations. As for the moving actions, there are four directions. If the target cell is an obstacle or the boundary, the agent is bounced back and remains in the previous cell.

The two robots in the Tiger-grid problem interact sparsely. The trigger point of interactions is when either robot opens the door with gold behind it. When it happens, both robots are reset to a random adjacent free cell of their previous positions as shown in Fig.~\ref{fig:exp-tiger-grid-reset}. In every other case, they act independently.

\subsubsection{Real-World Map Navigation} 
We select a set of 2D LIDAR maps picturing top-view layouts of real-world buildings from the Robotics 2D-Laser Datasets~\cite{laser-data}. We first down-sample them to appropriate sizes so that we can directly regard pixels as cells in Tiger grids. Then, we binarize the gray-scale pixels as obstacles and free space.

All the other settings of such realistic grids remain the same as in Tiger grids. Hence, the underlying configurations of \sipomdplitenettext{} models also remain the same, such that we can directly apply the model trained on Tiger grids to these real-world environments. We illustrate an example for this domain in Fig.~\ref{fig:exp-laser_maps}.

\subsection{Network Training}
\label{subsec:expr:training}

In both two categories of experiments, a task parameter $\bm{\theta}$ comprises both agents' grid maps, which illustrate the obstacle distribution and are typically identical, gold maps, which indicate the goal position for each agent and can be different, an initial belief of the subjective agent $i$ regarding the common states, and the indicator function for transition and reward interactions.

For each training task, we model agent $i$'s behaviors as an I-POMDP Lite and solve it with the MAQMDP algorithm, where agent $j$'s strategy is modeled and solved as a nested MDP. Thus, we obtain a series of simulated joint actions, $\langle\tau_{a_i},\tau_{a_j}\rangle$, and following observations, $\tau_{o_i}$, as the expert demonstration for this task. Then, we train the top-level architecture and the nested MDP module separately by minimizing the cross-entropy between the output and expert's action trajectories:

\begin{align}
  \label{eq:net-lossfunc}
  \begin{aligned}
    J(\bm{\theta}) = -\frac{1}{N}\frac{1}{|A_i|}\sum_{t=T}^{T+N-1}\sum_{a_i^t\in A_i}\log \Pr\nolimits_{\pi_i(\bm{\theta})}\bigg(a_i^t=\tau_{a_i}^t\bigg)
  \end{aligned}
\end{align}

In our experiments, the training set includes $100,000$ $6\times6$ and $50,000$ each of $10\times10$ and $12\times12$ Tiger grid tasks, with each task yielding an expert trajectory. We firstly train the \sipomdplitenettext{} on $6\times6$ grids and evaluate on multiple sizes of Tiger-grid tasks. The test set contains $500$ tasks for each grid size. Then, we further train the model on $10\times10$ and $12\times12$ grids and evaluate it on the LIDAR maps. We specify additional training details in Appendix~\ref{subsec:appendix:training} of the supplementary documentation.

\begin{table*}[t]
  \small
  \caption{We evaluate the \sipomdplitenettext{} by comparing the learned policy with the expert and model-free IA2C on multiple sizes of Tiger-grid tasks and four LIDAR maps. The network is trained on $6\times 6$ grids of the training set and evaluated on all $5$ sizes of grids in the test set. The trained model is then trained further on $10\times10$ and $12\times12$ grids and evaluated on the LIDAR maps. Overall, the \sipomdplitenettext{} has better performance than IA2C. It even outperforms the expert when transferring to larger environments.}
  \label{table:tab_results_tiger}
  \centering
  \setlength{\tabcolsep}{2.2pt}
  \begin{tabular}{lrrrrrrrrr}
    \toprule
    & \multicolumn{3}{c}{\textbf{I-POMDP Lite}} & \multicolumn{3}{c}{\textbf{\sipomdplitenettext{}}} & \multicolumn{3}{c}{\textbf{IA2C}}\\
    \cmidrule(lr){2-4}\cmidrule(lr){5-7}\cmidrule(lr){8-10}
    Env & Succ rate & F-open rate & Colli rate & Succ rate & F-open rate & Colli rate & Succ rate & F-open rate & Colli rate\\
    \midrule
    6$\times$6$^{\ast}$ & 0.858$\pm$0.003 & \textbf{0.208$\pm$0.002} & \textbf{0.059$\pm$0.002} & 0.851$\pm$0.003 & 0.235$\pm$0.003 & 0.070$\pm$0.003 & \textbf{0.940$\pm$0.367} & 0.410$\pm$0.245 & 0.220$\pm$0.392\\
    7$\times$7 & 0.812$\pm$0.003 & \textbf{0.192$\pm$0.002} & \textbf{0.056$\pm$0.002} & 0.804$\pm$0.003 & 0.215$\pm$0.003 & 0.085$\pm$0.002 & \textbf{0.820$\pm$0.240} & 0.460$\pm$0.382 & 0.240$\pm$0.284\\
    8$\times$8 & \textbf{0.820$\pm$0.003} & \textbf{0.188$\pm$0.002} & \textbf{0.066$\pm$0.003} & 0.790$\pm$0.004 & 0.204$\pm$0.003 & 0.098$\pm$0.004 & 0.690$\pm$0.424 & 0.440$\pm$0.316 & 0.210$\pm$0.258\\
    10$\times$10$^{\ast}$ & 0.712$\pm$0.004 & 0.150$\pm$0.002 & \textbf{0.076$\pm$0.003} &  \textbf{0.731$\pm$0.003} & 0.172$\pm$0.004 & 0.115$\pm$0.003 & 0.550$\pm$0.408 & 0.380$\pm$0.297 & 0.170$\pm$0.163\\
    12$\times$12$^{\ast}$ & 0.694$\pm$0.004 & \textbf{0.118$\pm$0.002} & \textbf{0.040$\pm$0.004} & \textbf{0.743$\pm$0.004} & \textbf{0.150$\pm$0.004} & 0.102$\pm$0.003 & 0.330$\pm$0.329 & 0.280$\pm$0.225 & 0.110$\pm$0.131\\
    \midrule
    ACES           & 0.830$\pm$0.023 & \textbf{0.098$\pm$0.028} & \textbf{0.064$\pm$0.034} & \textbf{0.842$\pm$0.029} & 0.112$\pm$0.030 & 0.080$\pm$0.036 & N/A & N/A & N/A\\
    Fr-camp      & \textbf{0.820}$\pm$0.017 & \textbf{0.125$\pm$}0.030 & \textbf{0.040$\pm$}0.031 & 0.803$\pm$0.025 & 0.150$\pm$0.036 & 0.048$\pm$0.034 & N/A & N/A & N/A\\
    Orebro        & 0.890$\pm$0.018 & \textbf{0.225$\pm$0.024} & \textbf{0.005$\pm$0.012} & \textbf{0.950$\pm$0.011} & 0.270$\pm$0.031 & 0.010$\pm$0.020 & N/A & N/A & N/A\\
    UW              & \textbf{0.940$\pm$0.021} & \textbf{0.172$\pm$0.022} & 0.032$\pm$0.012 & 0.910$\pm$0.017 & 0.188$\pm$0.041 & \textbf{0.020$\pm$0.015} & N/A & N/A & N/A\\
    \bottomrule
  \end{tabular}
\end{table*}

\subsection{Results and Discussion} 
\label{subsec:expr:results}
We show the results in Table~\ref{table:tab_results_tiger}, comparing our \sipomdplitenettext{} with an expert who solves the underlying I-POMDP Lite model and a novel model-free RL approach, IA2C~\cite{IA2C}, regarding their performance in multiple tasks. The mean and standard deviation of the statistics in the table are computed over $10$ rounds of $100$ simulations for each set of test tasks, of which the success rate is the proportion of the tasks where the agent opens the door with gold at least once and hits the obstacle at most once among all test tasks; the false-open rate is the proportion of the tasks in which agent ever opens the wrong door; and the collision rate is the proportion of the tasks ever collides the obstacle.

\paragraph{\sipomdplitenettext{} learns models that accurately approximate the underlying I-POMDP Lite framework.} This paragraph focuses on comparing the underlying I-POMDP Lite model and the NN analog learned by our network. To elucidate the learned transition and observation model well approximate the original one, we can directly compare the updated beliefs between the underlying framework and its NN analog. We visualize an instance of such belief update for eight consecutive steps. Due to the page limitation, we place the diagram in Appendix~\ref{subsec:appendix:model_exhibit}. It shows that for every time step when the two selects the same action, their corresponding beliefs are mostly consistent, validating that our network accurately learns the underlying model.

\paragraph{The \sipomdplitenettext{} possesses solid transfer capabilities, of which the trained model generalizes well to unseen environments without further training.} We train the \sipomdplitenettext{} on $6\times6$ tasks and evaluate the trained model on multiple sizes of tasks. The evaluating environments are unduplicated. On the other hand, the IA2C must be evaluated in the same environments where it is trained because it has no access to the map while training. Thus, the spatial location-dependent transitions it learns are restricted in certain environments, which cannot be transferred to another one. Looking into the table, we notice that the \sipomdplitenettext{}'s performance is inferior to the expert and IA2C for smaller environments but gradually overtaking them as environments increase. The expert's success rate degrades because the planning horizon is longer for larger environments, increasing its preference for the low-cost "stay and listen" over other mobility actions. As a result, the expert holds still in numerous tasks without reaching the goal even once. This also explains the anomalous descent of the expert's collision and false-open rate. The IA2C's performance draws a similar pattern but with a much sharper degradation. In contrast, the \sipomdplitenettext{} directly transfers its learned model to larger environments, so it has better exploring capability and reaches the gold more often, hence yielding a higher success rate but at the cost of higher collision and false-open rate. Furthermore, in experiments conducted for LIDAR maps~\footnote{We no longer experiment with IA2C on LIDAR maps because it learns almost nothing except "listen" unless we keep the starting point extremely close to the goal.}, we further train the previous model on $10\times10$ and $12\times12$. It shows that the \sipomdplitenettext{} policy performs comparably with the expert, which demonstrates sufficient transfer capabilities across various environments and state space.


\section{Conclusion}

This paper elucidates the \sipomdplitenettext{}, a deep recurrent policy network learning and solving the underlying I-POMDP Lite model supervised by the expert demonstration. It provides a simpler and more intuitive architecture than existing multiagent learning for planning approaches like IPOMDP-net, reducing the computational complexity.

We evaluate the \sipomdplitenettext{} on various tasks with different environment sizes, where the overall performance is comparable with the expert policies. In some tasks, the NN policy even outperforms the expert demonstration. Moreover, empirical results indicate that the NN policy generalizes well to unseen, larger, and more sophisticated tasks, showing solid transfer capabilities.

The future work includes meta-learning the recurrence number $K$ and learning the interaction indicator function by the network itself instead of being given as prior knowledge.



\clearpage

\appendix

\section{Appendix}
\label{appendix}

\subsection{Markovian Model Learning}
\label{subsec:appendix:model_exhibit}

We mention in the paper that the \sipomdplitenettext{} accurately learns the Markovian model of the underlying I-POMDP Lite framework. To prove the learned transition and observation model is sufficiently close to the original one, we can simulate the trained model and expert's policy, which is derived by solving the ground-truth I-POMDP Lite model, simultaneously, comparing the beliefs at each time-step.

\begin{figure}[h]
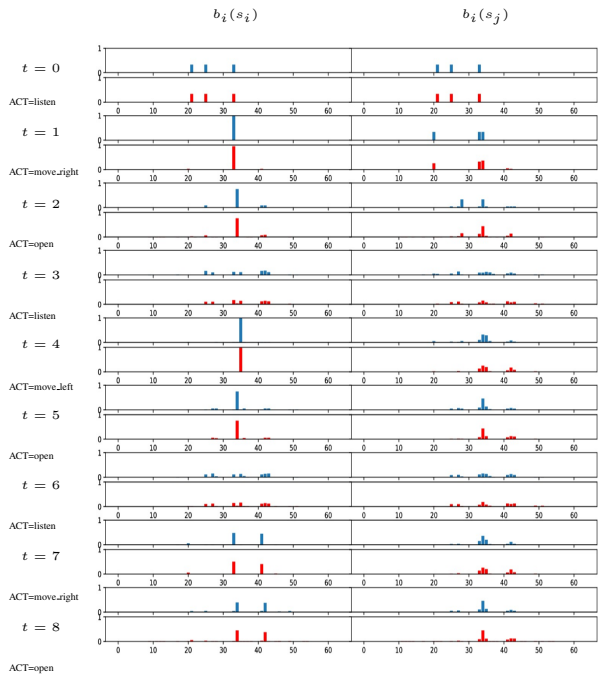

  \centering
  \includestandalone[width=0.5\textwidth]{belief_update_visual}
  \caption{The visualization of agent $i$'s belief update throughout a trajectory. The trajectory contains $8$ time steps. For each time step, the first row illustrates the belief given by the underlying I-POMDP Lite framework, while the second row depicts that of the IPOMDPLite-net. For a better intuition, we present the beliefs over $S_i$ and $S_j$ on the left and right sides, respectively.}
  \label{fig:b_update_visual}
\end{figure}

We visualize the updated belief for consecutive eight steps in Fig.~\ref{fig:b_update_visual}. Each row, consisting of two sub-rows, represents agent $i$'s belief updated by an action at a time step, where the blue bars are for the framework while red bars are for the trained model. To facilitate the visualization of beliefs, we draw $b_i(s)$ as $b_i(s_i)$ and $b_i(s_j)$, where $b_i(s_i) = \sum_{s_j}b_i(s)$ and $b_i(s_j) = \sum_{s_i}b_i(s)$. It shows that every two sub-rows in the diagram are mostly consistent, so our network accurately learns the underlying model from the expert.

\subsection{Ablation Study}
\label{subsec:appendix:ablation}
We carry out an ablation study on the $6\times6$ test set to demonstrate the importance of several \sipomdplitenettext{}'s components. We remove exactly one such component in every ablation experiment and train a model with the ablated architecture. We then study the role each component plays and to what extent it affects the network's performance.

\paragraph{Bayesian belief filter.} In this experiment, we remove the Bayesian filter that recursively updates the subjective agent's beliefs to investigate its necessity. Without the belief update, the network keeps taking in the initial belief and completely leaves the policy searching duty to the QMDP planning module. The ablation essentially eliminates the network's recurrent property, the critical factor that the network can account for the belief update in the I-POMDP Lite planning. We inspect the network's output action trajectories, finding that most of them are “stay and listen”, which explains the low collision rate. Therefore, the Bayesian belief filter is indispensable in our network.

\paragraph{Nested MDP planner.} In this experiment, we eliminate the nested MDP planner responsible for learning others' policies and predicting their actions to examine how considering others' intentions benefits the subjective agent in a multiagent system. The ablated network is essentially a QMDP-net. The second row of Table~\ref{table:tab_ablation} shows that the ablated version receives worse results for all test items. This is due to the single-agent planning being unaware of the potential interference caused by the other agent. For instance, when the other agent takes the lead to seize the gold and reset underlying states, the subjective agent will not realize it and listen in time; instead, it may continue according to the previous belief. Although the agent might choose to observe at some time steps by learning the expert demonstration and fortunately pull itself back on track, it is not aware of the other's behaviors; hence it will not substantially help the agent accurately locate itself in a multiagent system. As such, the nested MDP planner is indeed crucial.

\begin{table*}[t]
  \small
  \caption{Three ablation experiments for the 6$\times$6 Tiger-grid games and the impact on the success rate, false-open rate, and collision rate.}
  \label{table:tab_ablation}
  \centering
  \begin{tabular}{lccccc}
    \toprule
    Ablation & Succ rate & F-open rate & Colli rate\\
    \midrule
    \textbf{\sipomdplitenettext{}} & \textbf{0.851$\pm$0.003} &  \textbf{0.235$\pm$0.003} & \textbf{0.070$\pm$0.003}\\
    \midrule
    \sipomdplitenettext{} w/o belief update         & 0.001$\pm$0.000 & 0.001$\pm$0.000 & 0.115$\pm$0.000\\
    \sipomdplitenettext{} w/o nested MDP modeling   & 0.744$\pm$0.002 &  0.110$\pm$0.004 & 0.170$\pm$0.004\\
    \sipomdplitenettext{} w/o single-agent models & 0.063$\pm$0.001 & 0.001$\pm$0.000 & 0.875$\pm$0.001\\
    \bottomrule
  \end{tabular}
\end{table*}

\paragraph{Single-agent models trained for non-interactive situations.} In this experiment, we ablate all the single-agent models for both agents, including transition functions, i.e., $\mathfrak{t}_i(s_i,a_i,s_i^{\prime})$ and $\mathfrak{t}_j(s_j,a_j,s_j)$, and reward functions, i.e., $\mathfrak{r}_i(s_i,a_i)$ and $\mathfrak{r}_j(s_j,a_j)$, while relying on the multiagent models, i.e., $T_i(s,a,s^{\prime})$, $R_i(s,a)$, and $R_j(s,a)$, to take care of the training completely. In this case, the problem to solve is not necessarily under sparse interactions. However, according to the third row of Table~\ref{table:tab_ablation}, the trained multiagent models perform shockingly worse. This is due to the lack of guidance provided by the prior knowledge regarding the awareness of sparse interactions. With well-trained single-agent models that account for most of the inference, the multiagent models in training focus only on where interactions happen while ignoring the portion of non-interactive based on the attention mechanism of NNs. Therefore, the pre-trained single-agent models are imperative in our network for reasonably good performance concerning sparse interactions.

\subsection{Belief Update Module}
\label{subsec:appendix:bu}
We denote the input and output belief tensor by $\bm{B}_S$ and $\bm{B}_S^{\prime}$, respectively, where the subscript $S$ indicates the tensor shape, $|S|$. Both the two are beliefs of common states. In our paper, we assume that $S=S_i\times S_j$.

Consistent with the framework, the belief update module first executes the belief propagation with the joint action, $\langle a_i,a_j\rangle$. The initial step is to distinguish the interactive area from the non-interactive one within the state-action space. Specifically, we multiply $\bm{B}_S$ with $\bm{\X}_I^T$ and $1-bm{\X}_I^T$ element-wisely, where the dimensionality is broadcast from $|S|$ to $|S|\times|A|$. Thus, we decompose $\bm{B}_S$ into two complementary belief tensors. 

For the interactive part, we update $\bm{B}_S$ with $T_i(\cdot|\bm{\theta})$, a trainable variable approximating $T_i(s,a,s^{\prime})$. However, it is not necessarily of the shape $|S|\times|A|\times|S|$ in the network because we can choose diverse NN architectures to approximate the arithmetic dot product and summation depending on features of specific application scenarios. Based on the principle of learning as accurate a model as possible, we enforce the trained model to inherit properties of the underlying framework. In the case of $T_i(\cdot|\bm{\theta})$, we normalize it over the dimension of $s^{\prime}$ to ensure a valid probability distribution serving the Bayesian filtering. The network structure simulating the update with actions generally implements the below operation, where $\bm{v}_{a_i}$ and $\bm{v}_{a_j}$ are indexing vectors for $a_i$ and $a_j$.: 
\begin{align*}
    \begin{aligned}
     \bm{B}^{\prime}_{P_T^I} = \sum_s\sum_{a_i}\sum_{a_j}\bm{B}_{P_T^I}T_i(\cdot|\bm{\theta})\bm{v}_{a_i}\bm{v}_{a_j}
    \end{aligned}
\end{align*}

We update the belief for each $a\in A$, while we only concern about $\bm{B}_{P_T^I}^{\prime}$ corresponding to the input joint action $a=\langle a_i,a_j\rangle$.

For the non-interactive part, we execute the two-step belief propagation as shown in Fig.~\ref{fig:network_arch_beliefupdate}. Analogous to $T_i(\cdot|\bm{\theta})$, we define $\mathfrak{t}_i(\cdot|\bm{\theta}_i)$ and $\mathfrak{t}_j(\cdot|\bm{\theta}_j)$ to approximate the underlying $\mathfrak{t}_i(s_i,a_i,s_i^{\prime})$ and $\mathfrak{t}_j(s_j,a_j,s_j^{\prime})$. We consecutively conduct the belief propagation twice, computing the dot product between the non-interactive belief tensor and $\mathfrak{t}_j(\cdot|\bm{\theta}_j)$ and $\mathfrak{t}_i(\cdot|\bm{\theta}_i)$, respectively.

\ref{subsec:appendix:implement} demonstrates an instance of solving Tiger-grid problems with spatial locality, where we encode the belief propagation into a convolutional layer with $T_i(s,a,s^{\prime}|\bm{\theta})$ as the kernel.

\subsection{Value Iteration Solver}
\label{subsec:appendix:vi}
The value iteration solver implements the NN analog for multiagent MDP value iteration. Let us first consider the immediate reward. We define the NN counterparts of $R_i(s,a)$, $R_j(s,a)$, $\mathfrak{r}_i(s_i,a_i)$ and $\mathfrak{r}_j(s_j,a_j)$ as $R_i(s,a|\bm{\theta})$, $R_j(s,a|\bm{\theta})$, $\mathfrak{r}_i(s_i,a_i|\bm{\theta_i})$, and $\mathfrak{r}_j(s_j,a_j|\bm{\theta_j})$. In the \sipomdplitenettext{}, we condition $\mathfrak{r}_i(s_i,a_i|\bm{\theta_i})$ and $\mathfrak{r}_j(s_j,a_j|\bm{\theta_j})$ on their own environments in $\bm{\theta}$ while conditioning $R_i(s,a|\bm{\theta})$ and $R_j(s,a|\bm{\theta})$ on entire $\bm{\theta}$. For any given $s\in S$ and $a\in A$, there are $R_i(s,a)\in\mathbb{R}$ and $R_j(s,a)\in\mathbb{R}$, which is also true for $\mathfrak{r}_i(s_i,a_i)$ and $\mathfrak{r}_j(s_j,a_j)$, so we do not explicitly regulate the range of network's learning regarding the reward models.

The value iteration shared by the nested MDP and MAQMDP planner updates the action value function similarly to the manner that the belief tensor is updated, except that the order of the update is reverse.

After deriving the long-term expected reward, we add it to the immediate reward, i.e., $R_i(s,a|\bm{\theta})$. The network structure representing a recursion of the value iteration generally implements the following operation: 
\begin{align*}
    \begin{aligned}
     \bm{U}_S = &\max_{a_i\in A_i}\sum_{a_j\in A_j}\Bigg\{\sum_{s^{\prime}\in S}\bigg[\bm{U}^{\prime}_{S}T_i(\cdot|\bm{\theta})\bm{\X}_{P_T^I} + \bm{U}^{\prime}_{S}\mathfrak{t}_j(\cdot|\bm{\theta}_j)\\
     &\times\mathfrak{t}_i(\cdot|\bm{\theta}_i)(1-\bm{\X}_{P_T^I})\bigg] + R_i(s,a|\bm{\theta})\Bigg\}\bm{v}_{a_j}
    \end{aligned}
\end{align*}

\subsection{Nested MDP and MAQMDP Module}

The nested MDP planning module is an RNN with a hierarchical structure, which conforms to the setting of reasoning levels in the underlying nested MDP model. In this RNN, the hidden state is the policy, $\hat{\pi}_{i/j}^l$, output by each hierarchy. The most initial policy, which we denote by $\hat{\pi}^0$, is manually set to a uniform distribution over either $A_i$ or $A_j$, depending on the top reasoning level $L$, for all $s\in S$. We sample action from this initial policy and input it to the $0$th hierarchy of the planner to compute the level-$1$ policy. In the bottom hierarchy, $\hat{\pi}^0$ enters the embedded value iteration solver with the reward and transition functions. The value iteration solver outputs $Q^0(s,a)$ regarding joint actions. To get the policy for a specific agent, according to Eq.~\ref{eq:nested-mdp-vi}, we need to weight $Q^0(s,a)$ by the opponent's policy $\hat{\pi}^0$ and sum it over the dimension of the opponent's actions. Thus, we get the $Q$ values regarding only this agent's actions, on which we apply the Softmax operation along the dimension of actions to map $Q$ values to the policy, i.e., $\hat{\pi}^1$. Continuing with such procedures, we sample action from $\hat{\pi}^1$ and input it to the level-$1$ planner and compute $\hat{\pi}^2$ till it reaches the highest hierarchy.

\begin{figure}[h]
    \centering
    \includegraphics[width=0.5\textwidth]{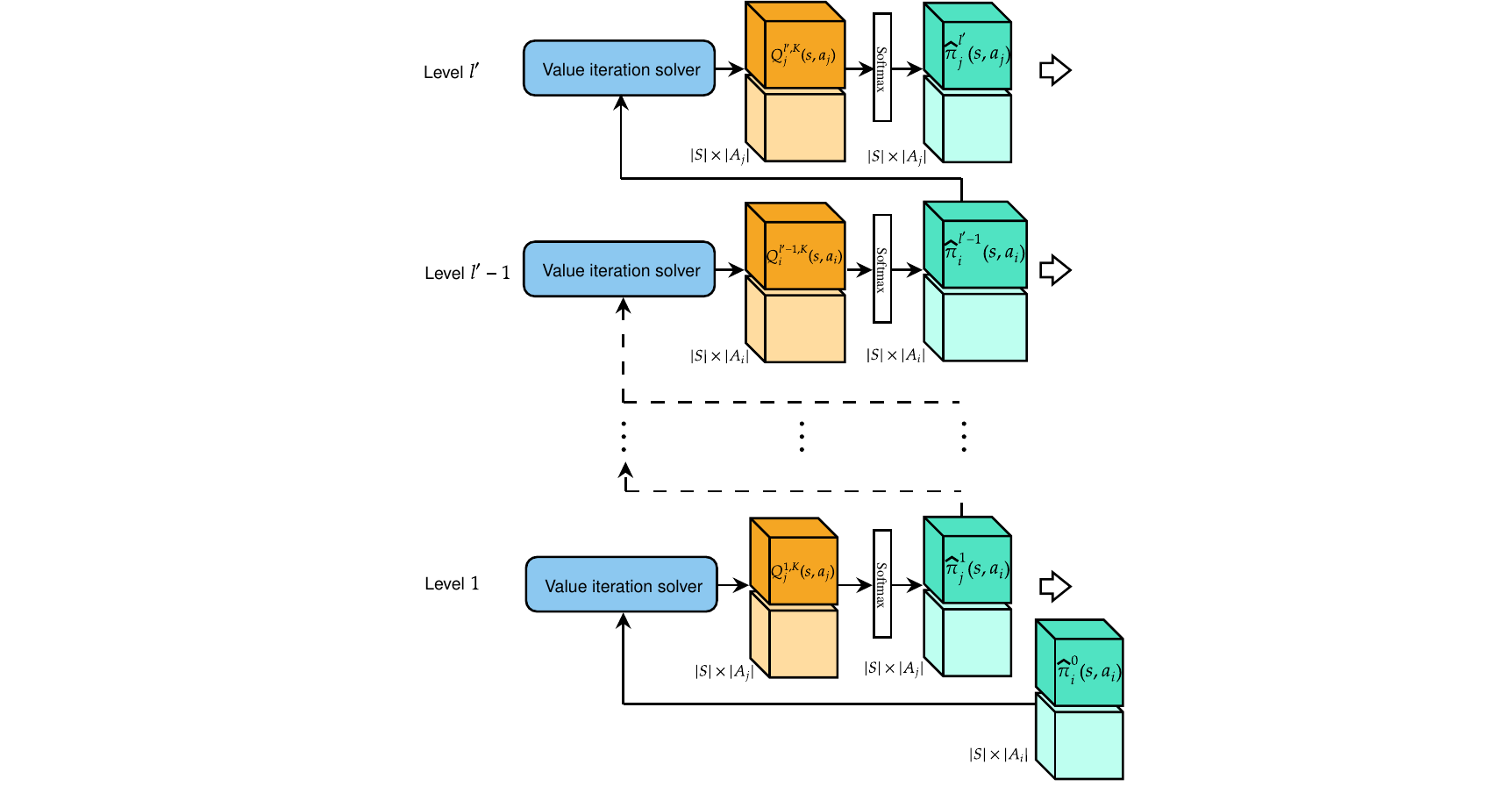}
    \caption{The hierarchical architecture of the nested MDP module. It generates the other agent's mixed strategy by alternately deriving the other agent's and its own policy at corresponding reasoning levels from the bottom up. The level-0 policy is a uniform distribution over the action space for whatever state.}
    \label{fig:network_arch_nested}
\end{figure}

Hierarchies that account for a certain agent's policy share the same input models and the network structure. As agent $i$ always reasons at the top level, $L$, where it applies the belief update and the QMDP solution, the highest level of the nested MDP module is $L-1$, where we solve $j$'s MDP to predict its policy and offer it to the level $L$ inference for $i$. Therefore, the $(L-1)$th hierarchy accepts $R_j$ and actions sampled from $\hat{\pi_i}^{L-1}$, while the $(L-2)$th hierarchy receives $R_i$ and actions from $\hat{\pi_j}^{L-2}$, and so on. All hierarchies share the convolutional filters representing transition functions, i.e., $\mathfrak{t}_i$, $\mathfrak{t}_j$, and $T_i$. we illustrate the general architecture of the nested MDP planning module in Fig.~\ref{fig:network_arch_nested}.

The MAQMDP planner's architecture can be regarded as a single level of the nested MDP hierarchical structure introduced with the updated belief. Instead of directly mapping the expected reward to the softmax policy, the MAQMDP planner first weight the output $Q$ function by the belief tensor from the belief update module, and the result is then converted to the policy.

\begin{table}[t]
  \small
  \caption{Reward setup for each agent under non-interactive situations. Under this situation, agents act as if interactions never occur and they are not impact by others. The \underline{FREESPACE} denotes such neighboring free cells, distinguishing by FREESPACE, which is the set of all free cells in the map.}
  \label{table:tab_non_interact_reward}
  \centering
  \setlength{\tabcolsep}{2.2pt}
  \begin{tabular}{lllr}
    \toprule
    Action & Current state & Target state & Reward\\
    \midrule
    LISTEN & $s\in$FREESPACE & $s^{\prime}=s$ & $-$0.2\\
    \midrule
    \multirow{2}{*}{MOVE} & $s\in$FREESPACE & $s^{\prime}\in$OBSTACLES & $-$10.0\\
    & $s\in$FREESPACE & $s^{\prime}\in$\underline{FREESPACE} & $-$0.5\\
    \midrule
    \multirow{2}{*}{OPEN} & $s=$GOLD & $s^{\prime}\in$\underline{FREESPACE} & $+$10.0\\
    & $s\in$FREESPACE$\setminus$GOLD & $s^{\prime}=s$ & $-$5.0\\
    \bottomrule
  \end{tabular}
\end{table}

\begin{figure}[h]
    \centering
    \includegraphics[width=0.5\textwidth]{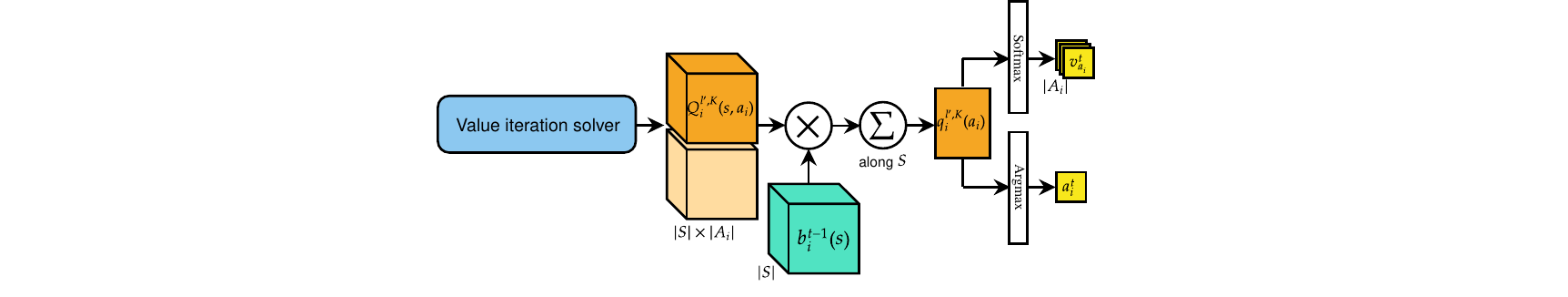}
    \caption{The MAQMDP planning module obtains the optimal policy for the subjective agent reasoning at the top level by weighting $Q$ values with the current belief and then selecting the best move based on the maximum value in the result.}
    \label{fig:network_arch_qmdp-planner}
\end{figure}

\subsection{Experimental Setup Details}
\label{subsec:appendix:setup}
We compress the observations into a tuple of four binary values corresponding to the existence of obstacles in its four cardinal neighboring cells.

In the single-agent setting, an agent must navigate itself to circumvent obstacles present randomly in cells and reach the cell where a pile of gold lies. Meanwhile, it only gets the gold by proactively executing the "open" action rather than simply being in the cell. When the agent chooses between staying and listening and opening the door in each cell, the problem becomes a Tiger problem. Table~\ref{table:tab_non_interact_reward} provides the single-agent rewards shared by all agents. The action of listening costs $-0.2$. Opening the door in a cell gains a reward of $10.0$ if the gold is behind the door; otherwise, it is cost $-5.0$. The action of moving integrates all individual Tiger problems happening in each cell and connects them with a navigation problem. Moving toward any direction has the identical cost of $-0.5$. Once the agent collides an obstacle, it receives a hurtful penalty of $-10.0$. The selection of gains and costs has to achieve a subtle balance. The agent should be neither too cautious and keep standing still nor too bold and moving recklessly without listening even if having collided with obstacles several times.

\begin{table}[t]
  \small
  \caption{Transition setup for each agent under non-interactive situations.}
  \label{table:tab_sa_trans}
  \centering
  \setlength{\tabcolsep}{2.2pt}
  \begin{tabular}{llll}
    \toprule
    Action & Current state & Target state & Next state\\
    \midrule
    LISTEN & $s\in$FREESPACE & $s^{\prime}=s$ & $\bm{s^{\prime}}=s^{\prime}=s$\\
    \midrule
    \multirow{2}{*}{MOVE} & $s\in$FREESPACE & $s^{\prime}\in$OBSTACLES & $\bm{s^{\prime}}=s$\\
    & $s\in$FREESPACE & $s^{\prime}\in$\underline{FREESPACE} & $\bm{s^{\prime}}=s^{\prime}$\\
    \midrule
    \multirow{2}{*}{OPEN} & $s=$GOLD & $s^{\prime}\in$\underline{FREESPACE} & $\bm{s^{\prime}}=s^{\prime}$\\
    & $s\in$FREESPACE$\setminus$GOLD & $s^{\prime}=s$ & $\bm{s^{\prime}}=s$\\
    \bottomrule
  \end{tabular}
\end{table}

\begin{table*}[t]
  \small
  \caption{Transitions for each agent under interactive situations. Based on the action setup of Tiger-grid problems, only when either agents opens the door might an interaction occur and the previous single-agent transitions alter. In short, when either agent successfully opens the door with gold, both will be reset to an adjacent free cell no matter what action the other one executes.}
  \label{table:tab_ma_trans}
  \centering
  \setlength{\tabcolsep}{2.2pt}
  \begin{tabular}{lllll}
    \toprule
    Action & Current state & Target state & SA next state & MA next state\\
    \midrule
    $\langle L,O\rangle$ & \FreeGold & $s_i^{\prime}=s_i$ & $\bm{s_i^{\prime}}=s_i^{\prime}=s_i$ & $\bm{s_i^{\prime}}\in$\underline{FREESPACE}\\
    \midrule
    $\langle M,O\rangle$ & \FreeGold & \primeObstFree & \BmNonprimePrime & \BmPrimeFree\\
    \midrule
    \multirow{2}{*}{$\langle O,O\rangle$} & \GoldGold & $s_i^{\prime}\in$\underline{FREESPACE} & $\bm{s_i^{\prime}}=s_i^{\prime}$ & $\bm{s_i^{\prime}}\in$\underline{FREESPACE}\\
    & \FreeMinusGoldGold & $s_i^{\prime}=s_i$ & $\bm{s_i^{\prime}}=s_i$ & $\bm{s_i^{\prime}}\in$\underline{FREESPACE}\\
    \bottomrule
  \end{tabular}
\end{table*} 

We make the Tiger-grid a multiagent system by imposing interference between involved agents. Specifically, both agents aim to seize the gold. Once an agent successfully achieves the goal, both of them are randomly relocated to a neighbor cell, specifically, a free cell in its cardinal and intercardinal directions, and the game continues. This design ensures interactions between the agents. By modeling the other agent's intentions, a self-interested agent keeps the awareness of where the other is and when it will open the door, which helps the agent determine its current situation and plan for the future. In the Tiger-grid, the uncertainty regarding dynamics is rooted in the reset of agents' locations. Hence, we keep the rest of the state transitions deterministic. We set the fault rate of observations to $0.01$ independently in each direction.

When either agent consecutively seizes the gold three times or the trajectory length exceeds the limitation, the trajectory terminates. Thus, we regard a trajectory as successful if the subjective agent seizes the gold at least once.

\subsection{Implementation Details}
\label{subsec:appendix:implement}
This section details the implementation of our work mainly from two aspects -- the generation of the training data, including the construction of task parameters and the creation of expert demonstrations, and the selection of specific network structures for some essential parts regarding a set of tasks represented by the Tiger-grid problems.

\subsubsection{Task Parameters And Expert Demonstration}
\label{subsubsec:expr:implement:theta-expertdemons}
Generally, a partial observable multiagent planning task requires the common state space of all involved agents as one of the environment-related prior knowledge and the initial belief over this state space shared by all agents as one of the agent-related prior knowledge.

In Tiger-grid problems, the grid map of an agent offers the basis for its observation function. Besides, the grid map and the goal map together determine its immediate rewards. Hence, when designing the network architecture for learning these underlying models, we explicitly condition them on specific components of $\bm{\theta}$. Hence, the task parameter of a Tiger-grid task comprises the navigation maps, goal maps, and the common initial belief for all agents. A particular component in our task parameter regarding the sparse-interaction setting is the interaction indicator functions for the given task. We deem them priors since learning them as hidden models through the neural network remains challenging at this time.

We initiate the construction of the Tiger-grid domain by randomly generating discrete grids. Each grid cell has a probability of $\Pr\sim U(0.1,0.3)$ to be an obstacle. The two agents share the identical action and observation space, where $A=\{a\in N|0\le a\le 5\}$ and $\Omega=\{0,1\}^4$. For each observation vector, an element of $1$ represents the agent observes that the adjacent cell in the corresponding direction is free space, while an element of $0$ means that it observes an obstacle in that cell. A common state of the domain consists of both agents' locations following a consistent order, while each agent takes grid cells as its private states, $S_i=S_j=\{s\in N|0\le s\le |M|^2\}$, where $M$ is the side length of the grid. Hence, the common state space is the Cartesian product of each agent's private state space, $S=\{(s_i, s_j)|0\le s_i\le |M|^2, 0\le s_j\le |M|^2\}$.

We then build underlying models for each grid environment. Since both agents' dynamics are local and spatial invariant, we use SciPy's sparse matrices to store the transition and reward functions. Thus, we can iteratively update beliefs through matrix multiplications and summations. Because an I-POMDP Lite approximately reduces to a POMDP and exactly solving a POMDP is expensive, we solve it with the more economic MA-QMDP and obtain a near-optimal policy. We implement the QMDP value iteration with the help of the MDP Toolbox, yielding the state utilities $V$ and action values $Q$.

We use the described planning scheme to generate expert trajectories. The input to the framework includes the initial physical state, terminal state, and belief over the initial state.  Both agents' initial actions are "stay and listen" by default. We save the output action and observation for each time step in an HDF5 database.

\subsubsection{Network Selection And Design for Problems with Spatial Locality}
\label{subsubsec:expr:implement:network}
For problems represented by the Tiger-grid, the states are defined in terms of spatial locations. If their actions can only lead to transitions from the current state to a subset of states within a certain range, we call such problems spatially local. 

We can employ CNNs into several vital parts of our network to solve such problems, including the belief propagation with actions and the long-term expected reward derivation in the value function, for an excellent approximation of their counterparts in the underlying framework.

The VIN and QMDP-net has demonstrated the viability of applying 2D convolutions to single-agent spatially local problems, where they approximate the belief propagation and the calculation for expected $Q$ values with a convolutional layer. The kernel of the layer plays a role as the transition function, which only captures the transition probabilities within the sliding window. The rationale of such approximation is that the operations with the transition function in the framework are consistent with the underlying logic of convolutions. Specifically, they all contain the dot product calculated by a set of multiplications and additions. A 2D convolution works as:
\begin{align*}
    \begin{aligned}
      \mathcal{O}[n,m] = (\mathcal{I}\ast\mathcal{K})[n,m] = \sum_q\sum_p\mathcal{K}[q,p]\mathcal{I}[n-q,m-p]
    \end{aligned}
\end{align*}
where $\mathcal{I}$, $\mathcal{O}$, and $\mathcal{K}$ are the input, output, and kernel (convolutional filters), respectively. The input and the output are of the same shape $n\times m$, and the kernel is of the shape $q\times p$. In single-agent problems, there is $|S|=n\times m$. For each $s=(n_s,m_s)\in S$, a valid target state $s^{\prime}=(n_{s^{\prime}},m_{s^{\prime}})$ after taking any action satisfies $n_{s^{\prime}}\in[n_s-r,n_s+r]$, $m_{s^{\prime}}\in[m_s-r,m_s+r]$, where $r$ is the step length of an action. Thus, if $p=2r+1$ and $q=2r+1$, and we regard $\mathcal{I}$ as the current belief $b(s)$, $\mathcal{O}$ as the updated belief $b^{\prime}(s^{\prime})$, and $\mathcal{K}$ as the transition for a specific action $a$, the equation becomes exactly the single-agent belief update:
\begin{align*}
    \begin{aligned}
      b^{\prime}(s^{\prime}) = \sum_s T_a(s,s^{\prime})b(s)
    \end{aligned}
\end{align*}

\begin{figure*}[tp]
    \centering
    \includegraphics[width=0.8\textwidth]{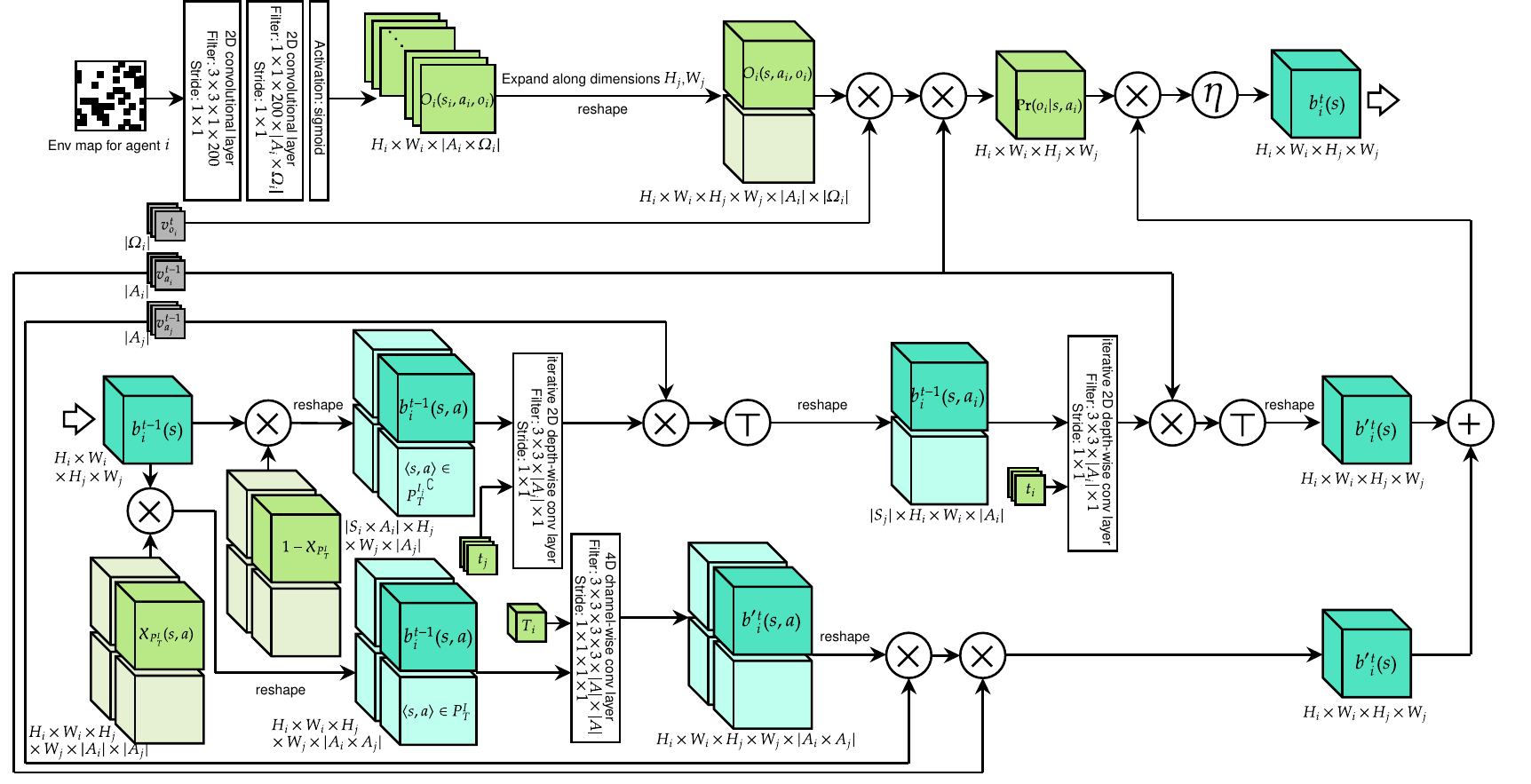}
    \caption{The specific architecture of \sipomdplitenettext{} that designed for addressing problems with spatial locality like Tiger-grid problems, where we approximate transition functions with kernels for convolutional layers. We also employ a CNN to learn the observation function.}
    \label{fig:network_arch_beliefupdate_spatloc}
\end{figure*}

The fact above illustrates the consistency between the belief update with actions and the convolution. The same applies to the computation for expected $Q$ values in the Bellman equation, where we swap the positions of $s$ and $s^{\prime}$.

We show that the applicability of using convolutions for approximation naturally carries over multiagent spatially local problems by expanding the dimensionality. Taking two-agent Tiger-grid problems as an example, we represent the common state space $S$ as a 4D tensor, the Cartesian product of the two 2D tensors representing $S_i$ and $S_j$. This representation maintains both agents' spatial localities as a whole.

Due to the overly computational intensity of high-dimensional convolutions, we only apply it for dealing with the interactions. As for non-interactive cases, we maintain the use of 2D convolutions. Hence, although the 4D kernel essentially represents the full transition function, we are only concerned about the part for interactions, not requiring equivalently accurate transitions to be learned for the non-interactive part, which benefits the training.

\begin{figure*}[t]
    \centering
    \includegraphics[width=0.8\textwidth]{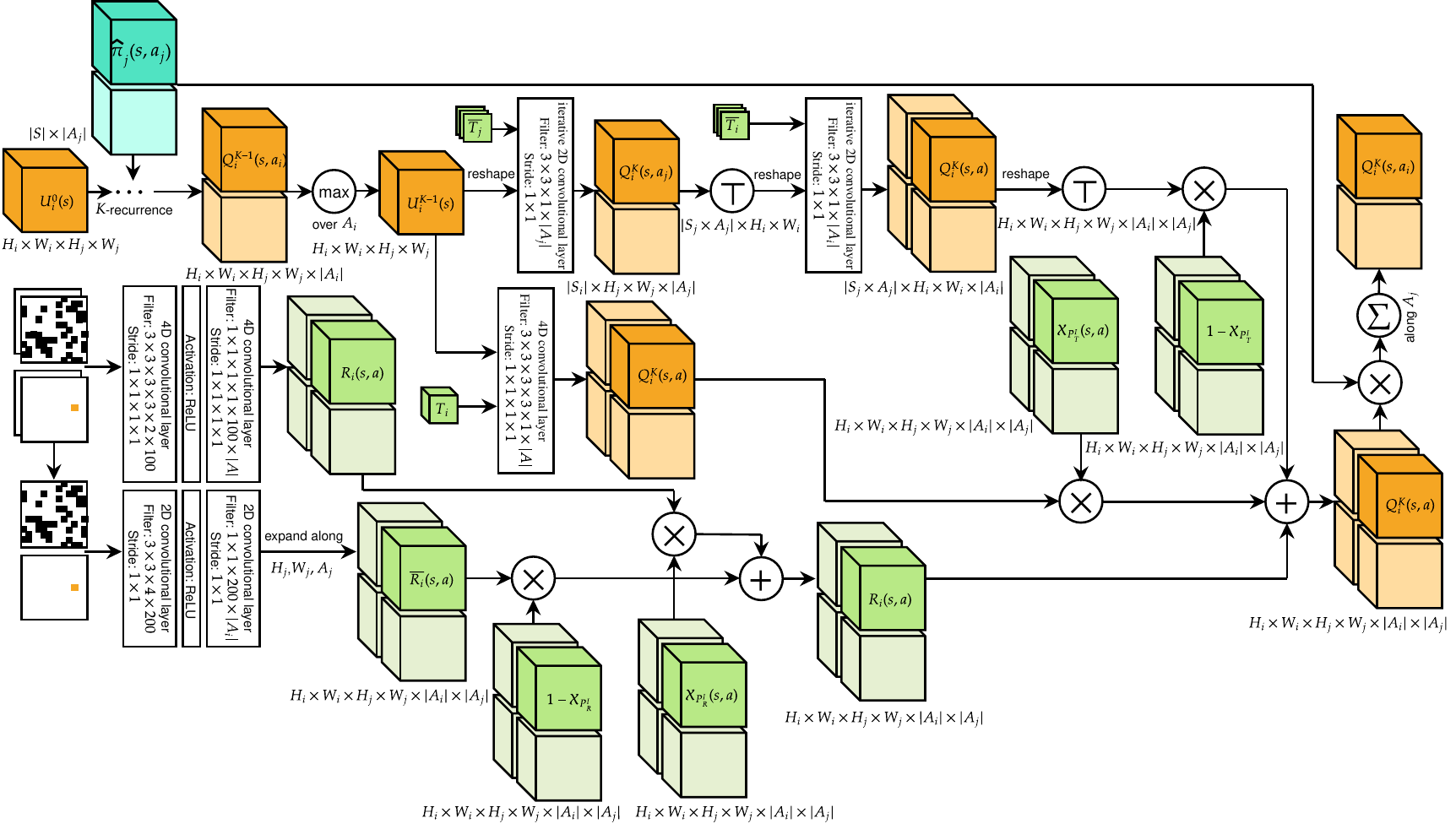}
    \caption{Value iteration solver of \sipomdplitenettext{} that to deal with spatially local problems such as Tiger-grid problems, where we approximate transition functions with kernels for convolutional layers and learn agents' reward functions with CNNs.}
    \label{fig:network_arch_vi-solver_spatloc}
\end{figure*}

In the rest of the subsection, we demonstrate the specific usage of convolutions for the belief update and value iteration in \sipomdplitenettext{} dealing with two-agent Tiger-grid problems and instantiate other important network structures that we mark in Fig.~\ref{fig:network_arch_beliefupdate_spatloc} and Fig.~\ref{fig:network_arch_vi-solver_spatloc}.

In two-agent Tiger-grid problems, each agent has a grid map in $\bm{\theta}$, which we define as their private state space, i.e., $S_i$ and $S_j$. Provided that each grid map is a 2D tensor, if agent $i$'s map is of the shape $H_I\times W_i$, and agent $j$'s is of $H_j\times W_j$, then we have $|S_i|=H_i\times W_i$ and $|S_j|=H_j\times W_j$. Hence, the common state space formed as a 4D tensor is of the shape $H_i\times W_i\times H_j\times W_j$, which is true for belief tensors.

As demonstrated in \ref{subsec:appendix:bu}, we first update the non-interactive part of the input belief tensor $\bm{B}_S$, corresponding to $\langle s,a\rangle\in {P_T^I}^{\complement}$. The resultant tensor after screening out the belief for ${P_T^I}^{\complement}$ by the indicator tensor $1-\bm{X_{P_T^I}}$, denoted by $\bm{B}_{{P_T^I}^{\complement}}$, is of the shape $H_i\times W_i\times H_j\times W_j\times A$, which we regard as a 4D image with $|A|$ channels. Provided that we should apply a two-step update here with the agents single-agent transition functions, $T_j$ and $T_i$, in order, and that the single-agent update is approximated by 2D convolutions, we need to first reshape $\bm{B}_{{P_T^I}^{\complement}}$ to match the 2D convolution with kernel $\mathcal{K}_{\mathfrak{t}_j}(s_j,a_j|\bm{\theta})$ and then transpose the result to match another 2D convolution with kernel $\mathcal{K}_{\mathfrak{t}_i}(s_i,a_i|\bm{\theta})$.

Continuing with this line, we address the two-step update for $\bm{B}_{{P_T^I}^{\complement}}$. When it comes to the update for $\bm{B}_{{P_T^I}}$, we initialize $\mathcal{K}_{T_i}(s,a|\bm{\theta})$ and apply it directly to the 4D convolution. After both updates, we select the channels corresponding to the given $a_i$ and $a_j$ and then sum them up, which yields the updated belief tensor $\bm{B}^{\prime}_S$.

When dealing with Tiger-grid problems, we learn agent $i$'s observation $O_i$ with a CNN, which captures the information of local environments for each $s_i\in S_i$ that matches the observations, specifically, the distribution of obstacles within a given range centered on $s_i$, and maps it to a valid representation of $\Pr(o_i|s_i,a_i)$ by forcing the weights on dimension $o_i\in\Omega_i$ to sum to $1$.

Next, to match $\bm{B}^{\prime}_S$ for the correction, we expand $O_i$ on dimensions $H_j$ and $W_j$ and tile the probabilities for the expanded dimensions. Finally, we get the fully updated belief tensor $\bm{B}_S$ by multiplying $\bm{B}^{\prime}_S$ and the expanded $O_i$ and normalizing the product over $S$.

We illustrate the details of convolutional layers in Fig.~\ref{fig:network_arch_beliefupdate_spatloc}.

The application of convolutions in the value iteration module is similar to that in the belief update. We still represent the two-agent transition function with a 4D kernel and represent the two single-agent transition functions with 2D kernels.

As for immediate rewards, we also learn them via CNNs. In Tiger-grids, if considering only the non-interactive situations, the immediate rewards for an agent depend on whether it opens the door in a Gold cell or not and whether it collides an obstacle or not. Hence, we condition the rewards on the grid map and goal map of the agent given in $\bm{\theta}$. We stack them together as a two-channel image and feed it into a CNN, where a kernel of the same shape as the transition function is applied. We learn single-agent reward functions, including $\mathfrak{r}_i(s_i,a_i)$ and $\mathfrak{r}_j(s_j,a_j)$, in this way. 

When it comes to the two-agent rewards, such as $R_i(s,a)$ and $R_j(s,a)$, where we must consider common states and joint actions, it is no longer sufficient to condition them on single-agent grid maps and goal maps. Instead, we merge the agents' respective grid maps and goal maps to compute the Cartesian product, hence getting a grid map and a goal map in the 4D space. Then, still stacking them together, we conduct the 4D convolution for several layers, mapping the two-channel image to a tensor that represents the two-agent reward function with $|A|=|A_i|\times|A_j|$ channels, each representing the reward of executing the corresponding joint action $a$ for all $s\in S$.

With all required immediate reward functions and the reward interaction indicators, we can get the true two-agent reward functions. We illustrate the details of CNNs in Fig.~\ref{fig:network_arch_vi-solver_spatloc}.

\subsection{Training Details}
\label{subsec:appendix:training}

\subsubsection{Training with Expert Demonstration}
\label{subsubsec:appendix:training:imit-learning}
We train the \sipomdplitenettext{} with the supervision of expert demonstration. Compared to RL methods, this training approach presents higher data efficiency. Rather than beginning with millions of randomly generated transitions and their corresponding rewards, we require much fewer expert action trajectories representing the near optimal policy we aim to learn. Hence, our goal is to minimize the cross entropy between action trajectories given by the expert demonstration and by our network.

We train the \sipomdplitenettext{} asynchronously. Specifically, we separate the whole network architecture into the outer part, including the belief update and MAQMDP planning module, and the inner part, i.e., the nested MDP module. When training the former, the network receives a joint action trajectory and an observation trajectory, and output an action trajectory regarding the subjective agent. Then, it learns by minimizing the cross entropy between this trajectory and the expert's demonstrated action trajectory. When training the latter, the network only accepts a joint action trajectory. If there is only one nested reasoning level, i.e., agent $j$ reasons at level $0$, then the output of the nested MDP module is an action trajectory of agent $j$. We provide the expert demonstration for agent $j$ this time and minimize the cross entropy between the two trajectories.

In this way, we train the two major \sipomdplitenettext{} components of respectively. When it comes to the evaluation, we first apply the well-trained nested MDP planner to get agent $j$'s action at each time step. Next, we join it with agent $i$'s action and feed it into the top-level belief update and MAQMDP planning module. The outputs of both parts then participate in the next step's reasoning and planning.

We train a policy using demonstrated trajectories from 10,000 random $6\times6$ environments, $10$ trajectories with different initial locations, initial beliefs, and gold positions for both agents from each environment. We then evaluate the trained model on separate sets of random environments with gradually larger sizes, each with $500$ environments. Finally, we repeat the evaluation for the trained policy on each task $100$ times to compute the average values mitigating the impact of stochasticity during the simulation.

In \ref{subsec:appendix:implement}, we have shown the procedures of creating expert trajectories and storing them in the database. We then introduce how these trajectories are processed for mini-batch training and imitation learning.

Since we only choose successful expert trajectories to be our demonstration, we first filter out failed ones. Then, we produce mini-batches for backpropagation through time (BPTT), which is widely used for training recurrent NN (RNN). We break down a full trajectory into several sub-trajectories with the length equal to the backpropagation step size, each of which is wrapped by a block. We set the step size to $5$ for $6\times6$ and $7\times7$ Tiger-grids and to $8$ when training the model further on $8\times8$, $10\times10$, and $12\times12$ grids. Provided with the total number of blocks needed, we get the number of steps for each epoch by dividing it using the mini-batch size. We set the mini-batch size to $100$ for all of our experiments. Next, we joint trajectories end to end in blocks for all mini-batches. Once a trajectory terminates, we pad it until reaching the block limitation. New trajectories begin from the next block even if other trajectories in the batch have not terminated yet. We still need to distinguish if a sub-trajectory is the start of its original trajectory. This determines whether we should assign the initial belief to it or not. As such, we complete creating batch samples. The final step of yielding training data is to extract actions and observations from stored steps of all trajectories and specify any pair of consecutive actions as the input and target (or label) action, respectively. Hence, we construct the mapping from an input action-observation pair to the corresponding next action in the demonstration.

To apply the imitation learning, we intuitively define the loss as the cross-entropy between the network's predicted actions and the demonstrated actions along the expert trajectories. 

We apply RMSProp optimizer with $0.9$ decay rate and $0$ momentum. The learning rate is $1\times10^{-3}$ for training from scratch and $1\times10^{-5}$ for further training a trained model with new data. We combine early stopping with patience and exponential learning rate decay for the adaptive gradient descent. Specifically, we set the initial patience to $30$ epochs and the rest to $10$ epochs; and we perform $20$ iterations of learning rate decay in total. It means that we do not decay the learning rate at first until the loss does not decrease for $40$ consecutive epochs on the validation set. Subsequently, we decay the current learning rate by $0.9$ if the loss does not decrease for $15$ epochs with it. We set the ratio of the training set and validation set to $9:1$.

Another essential hyperparameter for our network is the number of value iterations. In the underlying planning framework, the iteration terminates when it converges to the true state utilities $\left\Vert U_{i+1}-U_i\right\Vert<\epsilon(1-\gamma)/\gamma$, where $U_i$ and $U_{i+1}$ are utilities of the $i$th and $(i+1)$th iteration, $\epsilon$ is the maximum error acceptable for convergence, and $\gamma$ is the discounted factor. 

However, in our network, we execute the value iteration for a specific number of steps given by the hyperparameter $K$. We train the network with $6\times6$ Tiger-grid domain with randomly generated environments and then directly apply the trained model to tasks with larger environments, i.e., larger state space. We select $K$ for a set of tasks based on empirical trials. We first select a benchmark value for it, where $K=4N$. Then, we search for the best $K$ within the range of $[4N-10,4N+10]$ in units of $5$ by evaluating the policy trained with each value of $K$. We pick the one with the best performance to further evaluate the policy on larger tasks. We list the values selected for $K$ regarding each $N$ in \ref{table:tab_hyperparams}.

Table~\ref{table:tab_hyperparams} presents all the important hyperparameters that we discussed above.

\begin{table}[htbp]
  \small
  \caption{Essential hyperparameters.}
  \label{table:tab_hyperparams}
  \centering
  \begin{tabular}{lr}
    \toprule
    Hyperparameter & Argument\\
    \midrule
    Step size ($N=6$)             & 5\\
    Step size ($N={10,12}$)       & 6\\
    Mini-batch size               & 50\\
    Initial learning rate (from scratch) & 1$\times$10$^{-3}$\\
    Initial learning rate (re-training)  & 1$\times$10$^{-4}$\\
    Maximum epochs                & 1$\times$10$^3$\\
    Train-valid ratio             & 9 : 1\\
    \midrule
    $K$ ($N=6$)                   & 24\\
    $K$ ($N=7$)                   & 30\\
    $K$ ($N=8$)                   & 32\\
    $K$ ($N=10$)                  & 40\\
    $K$ ($N=12$)                  & 50\\
    \bottomrule
  \end{tabular}
\end{table}

\subsubsection{Transferring Pre-learned Knowledge}
\label{subsubsec:appendix:training:pretrain-transfer}
We train the network to learn transition functions including $T_i(s_i,a_i)$, $T_j(s_j,a_j)$, and $T_i(s,a)$, reward functions including $R_i(s_i,a_i)$, $R_i(s,a)$, $R_j(s_j,a_j)$, and $R_j(s,a)$, and the observation function $O_i(s,a_i,o_i)$. Learning these models from scratch simultaneously via backpropagation is prohibitively difficult. We alleviate this dilemma by dividing the training procedure into multiple steps, learning part of the models at first and then using them as priors to continue training the remaining models. Given that agents mostly follow their single-agent models in tasks, we can first learn them based on single-agent demonstrations. We actually train several distinct QMDP-nets and obtain $T_i(s_i,a_i)$, $T_j(s_j,a_j)$, $R_i(s_i,a_i)$, $R_j(s_j,a_j)$, and $O_i(s_i,a_i,o_i)$. After that, we freeze the weights of these learned models while focusing on training the multiagent models for interactions. In this way, we guide the network training in the right direction, which is another sense of effective use of a priori knowledge.


\begin{thebibliography}{}

\bibitem[Cyrill et~al., 2014]{laser-data}
Cyrill, S., Giorgio, G., Dirk, H., Henrik, A., Per, L., Tom, D., and Patrick,
  B. (2014).
\newblock Pre-2014 robotics 2d-laser datasets.
\newblock \url{http://www.ipb.uni-bonn.de/datasets/}.

\bibitem[Doshi and Gmytrasiewicz,
  2009]{decision-theoretic-framework-solver:apprx-I-PF}
Doshi, P. and Gmytrasiewicz, P.~J. (2009).
\newblock Monte carlo sampling methods for approximating interactive pomdps.
\newblock {\em Journal of Artificial Intelligence Research}, 34:297--337.

\bibitem[Doshi and Perez,
  2008]{decision-theoretic-framework-solver:apprx-I-PBVI}
Doshi, P. and Perez, D. (2008).
\newblock Generalized point based value iteration for interactive pomdps.
\newblock In {\em AAAI}, pages 63--68.

\bibitem[Han and Gmytrasiewicz, 2019]{neural-learning-planning:IPOMDP-net}
Han, Y. and Gmytrasiewicz, P. (2019).
\newblock Ipomdp-net: A deep neural network for partially observable
  multi-agent planning using interactive pomdps.
\newblock In {\em Proceedings of the AAAI Conference on Artificial
  Intelligence}, volume~33, pages 6062--6069.

\bibitem[Hansen et~al., 2004]{game-theoretic-framework:DP-POSGs}
Hansen, E.~A., Bernstein, D.~S., and Zilberstein, S. (2004).
\newblock Dynamic programming for partially observable stochastic games.
\newblock In {\em AAAI}, volume~4, pages 709--715.

\bibitem[He et~al., 2021]{IA2C}
He, K., Banerjee, B., and Doshi, P. (2021).
\newblock Cooperative-competitive reinforcement learning with history-dependent
  rewards.
\newblock In {\em Proceedings of the 20th International Conference on
  Autonomous Agents and MultiAgent Systems}, pages 602--610.

\bibitem[Hoang and Low, 2013]{decision-theoretic-framework:I-POMDP-Lite}
Hoang, T.~N. and Low, K.~H. (2013).
\newblock Interactive pomdp lite: Towards practical planning to predict and
  exploit intentions for interacting with self-interested agents.
\newblock In {\em Proceedings of the Twenty-Third International Joint
  Conference on Artificial Intelligence}, IJCAI '13, page 2298–2305. AAAI
  Press.

\bibitem[Hu et~al., 1998]{game-theoretic-framework:MARL}
Hu, J., Wellman, M.~P., et~al. (1998).
\newblock Multiagent reinforcement learning: theoretical framework and an
  algorithm.
\newblock In {\em ICML}, volume~98, pages 242--250. Citeseer.

\bibitem[Karkus et~al., 2017]{neural-learning-planning:QMDP-net}
Karkus, P., Hsu, D., and Lee, W. (2017).
\newblock Qmdp-net: Deep learning for planning under partial observability.

\bibitem[Littman et~al., 1995]{qmdp-1}
Littman, M.~L., Cassandra, A.~R., and Kaelbling, L.~P. (1995).
\newblock Learning policies for partially observable environments: Scaling up.
\newblock In {\em ICML}.

\bibitem[Melo and Veloso, 2011]{Dec-SIMDP}
Melo, F.~S. and Veloso, M. (2011).
\newblock Decentralized mdps with sparse interactions.
\newblock {\em Artificial Intelligence}, 175(11):1757--1789.

\bibitem[Melo and Veloso, 2013]{heuristic-Dec-SIMDP}
Melo, F.~S. and Veloso, M. (2013).
\newblock Heuristic planning for decentralized mdps with sparse interactions.
\newblock In {\em Distributed Autonomous Robotic Systems}, pages 329--343.
  Springer.

\bibitem[Ng et~al., 2010]{critique:I-POMDP}
Ng, B., Meyers, C., Boakye, K., and Nitao, J.~J. (2010).
\newblock Towards applying interactive pomdps to real-world adversary modeling.
\newblock In {\em IAAI}.

\bibitem[Pineau, 2004]{qmdp-2}
Pineau, J. (2004).
\newblock {\em Tractable planning under uncertainty: exploiting structure}.
\newblock Carnegie Mellon University.

\bibitem[Tamar et~al., 2016]{neural-learning-planning:VIN}
Tamar, A., WU, Y., Thomas, G., Levine, S., and Abbeel, P. (2016).
\newblock Value iteration networks.
\newblock In Lee, D., Sugiyama, M., Luxburg, U., Guyon, I., and Garnett, R.,
  editors, {\em Advances in Neural Information Processing Systems}, volume~29.
  Curran Associates, Inc.

\bibitem[Vinyals et~al., 2019]{game-theoretic-DMARL-grand-SC2}
Vinyals, O., Babuschkin, I., Czarnecki, W.~M., Mathieu, M., Dudzik, A., Chung,
  J., Choi, D.~H., Powell, R., Ewalds, T., Georgiev, P., et~al. (2019).
\newblock Grandmaster level in starcraft ii using multi-agent reinforcement
  learning.
\newblock {\em Nature}, 575(7782):350--354.

\bibitem[Yang et~al., 2018]{game-theoretic-DMARL-meanfield}
Yang, Y., Luo, R., Li, M., Zhou, M., Zhang, W., and Wang, J. (2018).
\newblock Mean field multi-agent reinforcement learning.
\newblock In {\em International Conference on Machine Learning}, pages
  5571--5580. PMLR.

\end{thebibliography}


\clearpage
\end{document}